\documentclass[twocolumn]{aastex631}

\usepackage{amsmath}
\usepackage{booktabs}
\usepackage{tabularx}
\usepackage{threeparttable}
\usepackage{tablefootnote}
\usepackage[title]{appendix}
\usepackage{float}

\shorttitle{Energy Budget of WASP-121\,$\rm{b}$ from JWST/NIRISS Phase Curve}
\shortauthors{Splinter et al.}

\begin{document}

\title{Precise Constraints on the Energy Budget of WASP-121\,b from its JWST NIRISS/SOSS Phase Curve}

\author[0000-0001-9987-467X]{Jared Splinter}
\affiliation{Trottier Space Institute at McGill, 3550 rue University, Montr\'eal, QC H3A 2A7, Canada}
\affiliation{Department of Earth and Planetary Sciences, McGill University, 3450 rue University, Montr\'eal, QC H3A OE8, Canada}

\author[0000-0002-2195-735X]{Louis-Philippe Coulombe}
\affiliation{Institut Trottier de recherche sur les exoplan\`etes, Département de Physique, Universit\'e de Montr\'eal, Montr\'eal, Québec, Canada}

\author[0000-0001-6569-3731]{Robert C. Frazier}
\affiliation{Department of Astronomy, University of Michigan, 1085 S. University Ave, Ann Arbor, MI 48109, USA} 

\author[0000-0001-6129-5699]{Nicolas B. Cowan}
\affiliation{Trottier Space Institute at McGill, 3550 rue University, Montr\'eal, QC H3A 2A7, Canada}
\affiliation{Department of Earth and Planetary Sciences, McGill University, 3450 rue University, Montr\'eal, QC H3A OE8, Canada}
\affiliation{Department of Physics, McGill University, 3600 rue University, Montr\'eal, QC H3A 2T8, Canada}

\author[0000-0003-3963-9672]{Emily Rauscher}
\affiliation{Department of Astronomy, University of Michigan, 1085 S. University Ave, Ann Arbor, MI 48109, USA} 

\author[0000-0003-4987-6591]{Lisa Dang}
\affiliation{Institut Trottier de recherche sur les exoplan\`etes, Département de Physique, Universit\'e de Montr\'eal, Montr\'eal, Québec, Canada}
\affiliation{Department of Physics and Astronomy, University of Waterloo, 200 University W, Waterloo, Ontario, Canada, N2L 3G1}
\affiliation{Waterloo Centre for Astrophysics, University of Waterloo, 200 University W, Waterloo, Ontario, Canada, N2L 3G1}

\author[0000-0002-3328-1203]{Michael Radica}
\affiliation{Department of Astronomy \& Astrophysics, University of Chicago, 5640 South Ellis Avenue, Chicago, IL 60637, USA}
\author[0009-0002-9657-6874]{Sean Collins}  
\affiliation{Department of Physics and Astronomy, University of British Columbia, Vancouver, BC, Canada}
\affiliation{Trottier Space Institute at McGill, 3550 rue University, Montr\'eal, QC H3A 2A7, Canada}
\affiliation{Institut Trottier de recherche sur les exoplan\`etes, Département de Physique, Universit\'e de Montr\'eal, Montr\'eal, Québec, Canada}

\author[0000-0002-8573-805X]{Stefan Pelletier}
\affiliation{Observatoire astronomique de l'Université de Genève, 51 chemin Pegasi 1290 Versoix, Switzerland}

\author[0000-0002-1199-9759]{Romain Allart}
\altaffiliation{SNSF Postdoctoral Fellow}
\affiliation{Institut Trottier de recherche sur les exoplan\`etes, Département de Physique, Universit\'e de Montr\'eal, Montr\'eal, Québec, Canada}

\author[0000-0003-4816-3469]{Ryan J. MacDonald}
\affiliation{Department of Astronomy, University of Michigan, 1085 S. University Ave, Ann Arbor, MI 48109, USA} 

\author[0000-0002-6780-4252]{David Lafreni\`{e}re}
\affiliation{Institut Trottier de recherche sur les exoplan\`etes, Département de Physique, Universit\'e de Montr\'eal, Montr\'eal, Québec, Canada}

\author[0000-0003-0475-9375]{Lo\"{i}c Albert}
\affiliation{Institut Trottier de recherche sur les exoplan\`etes, Département de Physique, Universit\'e de Montr\'eal, Montr\'eal, Québec, Canada}

\author[0000-0001-5578-1498]{Bj\"orn Benneke}
\affiliation{Department of Earth, Planetary, and Space Sciences, University of California, Los Angeles, CA USA}
\affiliation{Institut Trottier de recherche sur les exoplan\`etes, Département de Physique, Universit\'e de Montr\'eal, Montr\'eal, Québec, Canada}

\author[0000-0001-5485-4675]{Ren\'{e} Doyon}
\affiliation{Institut Trottier de recherche sur les exoplan\`etes, Département de Physique, Universit\'e de Montr\'eal, Montr\'eal, Québec, Canada}

\author[0000-0001-5349-6853]{Ray Jaywardhana}
\affiliation{Department of Physics and Astronomy, Johns Hopkins University, Baltimore, MD 21218, USA}

\author[0000-0002-6773-459X]{Doug Johnstone}
\affiliation{NRC Herzberg Astronomy and Astrophysics, 5071 West Saanich Rd, Victoria, BC, V9E 2E7, Canada }
\affiliation{Department of Physics and Astronomy, University of Victoria, Victoria, BC, V8P 5C2, Canada}

\author[0000-0003-2310-9415]{Vigneshwaran Krishnamurthy}
\affiliation{Trottier Space Institute at McGill, 3550 rue University, Montr\'eal, QC H3A 2A7, Canada}
\affiliation{Department of Physics, McGill University, 3600 rue University, Montr\'eal, QC H3A 2T8, Canada}

\author[0000-0002-2875-917X]{Caroline Piaulet-Ghorayeb}
\altaffiliation{E. Margaret Burbridge Postdoctoral Fellow}
\affiliation{Department of Astronomy \& Astrophysics, University of Chicago, 5640 South Ellis Avenue, Chicago, IL 60637, USA}

\author[0000-0002-0436-1802]{Lisa Kaltenegger}
\affiliation{Department of Astronomy and Carl Sagan Institute, Cornell University, Ithaca, NY 14853, USA}

\author[0000-0003-1227-3084]{Michael R. Meyer}
\affiliation{Department of Astronomy, University of Michigan, 1085 S. University Ave, Ann Arbor, MI 48109, USA} 

\author[0000-0003-4844-9838]{Jake Taylor}
\affiliation{Department of Physics, University of Oxford, Parks Rd, Oxford OX1 3PU, UK}

\author[0000-0001-7836-1787]{Jake D. Turner}
\affiliation{Department of Astronomy and Carl Sagan Institute, Cornell University, Ithaca, NY 14853, USA}

\begin{abstract}
Ultra-hot Jupiters exhibit day-to-night temperature contrasts upwards of 1000\,K due to competing effects of strong winds, short radiative timescales, magnetic drag, and H$_2$ dissociation/recombination. Spectroscopic phase curves provide critical insights into these processes by mapping temperature distributions and constraining the planet’s energy budget across different pressure levels. Here, we present the first NIRISS/SOSS phase curve of an ultra-hot Jupiter, WASP-121\,b. The instrument's bandpass [0.6--2.85 $\mu$m] captures an estimated 50--83\% of the planet's bolometric flux, depending on orbital phase, allowing for unprecedented constraints on the planet's global energy budget; previous measurements with HST/WFC3 and JWST/NIRSpec/G395H captured roughly 20\% of the planetary flux. Accounting for the unobserved regions of the spectrum, we estimate effective day and nightside temperatures of $T_{\rm day} = 2717 \pm 17$K and $T_{\rm night} = 1562^{+18}_{-19}$K corresponding to a Bond albedo of $A_\mathrm{B} = 0.277\, \pm0.016$ and a heat recirculation efficiency of $\epsilon = 0.246\, \pm 0.014$. Matching the phase-dependent effective temperature with energy balance models yields a similar Bond albedo of 0.3 and a mixed layer pressure of 1\,bar consistent with photospheric pressures, but unexpectedly slow winds of 0.2 km/s, indicative of inefficient heat redistribution. The shorter optical wavelengths of the NIRISS/SOSS Order 2 yield a geometric albedo of $A_\mathrm{g} = 0.093^{+ 0.029}_{- 0.027}$ (3$\sigma$ upper limit of 0.175), reinforcing the unexplained trend of hot Jupiters exhibiting larger Bond albedos than geometric albedos. We also detect near-zero phase curve offsets for wavelengths above 1.5\,$\mu$m, consistent with inefficient heat transport, while shorter wavelengths potentially sensitive to reflected light show eastward offsets. 
\end{abstract}

\section{Introduction}

Ultra-hot Jupiters represent some of the most extreme natural laboratories for exploring atmospheric physics, chemistry, and dynamics, particularly the mechanisms governing energy transport. These planets are generally tidally locked due to their short orbital periods, resulting in a permanent dayside facing the host star and a nightside facing toward space, creating large day-night temperature and pressure gradients which drive vigorous atmospheric circulation \citep{Showman-Guillot2002, Heng-Showman2015}. WASP-121\,b is one of the most thoroughly studied ultra-hot Jupiters \citep{Delrez2016-discovery}. This highly irradiated gas giant ($R_\mathrm{p}\approx1.75 R_{\rm Jup}$, $M_\mathrm{p}\approx1.16 M_{\rm Jup}$) orbits its F6V-type host star in only 1.27 days, placing it among the largest and hottest ($T_{\rm eq} = 2358 \pm 52$\,K) exoplanets known \citep{Delrez2016-discovery, Bourrier2020}. Due to its inflated size and high temperatures, WASP-121\,b is an ideal target for atmospheric studies, enabling high signal-to-noise observations at near-infrared wavelengths.

The planet’s atmosphere has been studied extensively through transit observations \citep{Evans2018, Ben-Yami2018, Gibson2020, Sing2019, Hoeijmakers2020, Evans2016, Gapp2025-NIRSpec-transmission} and secondary eclipse measurements \citep{Evans2017, Mikal-Evans2019, Mikal-Evans2020}, revealing an atmosphere rich with atomic and molecular species. While transmission spectra probe the upper layers near the terminator and eclipse observations cover the full dayside, each captures only isolated portions of a complex, dynamic atmosphere. Phase curve measurements complement these techniques by revealing how heat is transported across the entire planet, enabling a more complete understanding of atmospheric dynamics.

Exoplanet phase curves map global temperature distributions and atmospheric compositions across a planet’s day and night sides \citep{Phasecurve-book}. Spectroscopic phase curves offer an advantage over photometric curves by probing a broader range of atmospheric pressures, providing pressure-dependent insights into advective and radiative timescales \citep{Stevenson2014}. To this end, WASP-121\,b has also had a number of photometric and spectroscopic phase curve observations with instruments such as TESS \citep{Bourrier2020,Daylan2021}, \textit{Spitzer} \citep{Morello2023}, \textit{Hubble} \citep{Mikal-Evans2022}, and \textit{JWST} NIRSpec \citep{Mikal-Evans2023, Evans-Soma2025-NIRSpec-PhaseCurve}. These observations have revealed a low geometric albedo and strong phase-curve modulations, marked by pronounced day-night temperature contrasts and minimal phase offsets, suggestive of a low Bond albedo and inefficient heat redistribution within the atmosphere.

The efficiency of horizontal energy transport on synchronously rotating exoplanets is roughly predicted by the ratio of radiative to advective timescales, which in turn is closely tied to the planet's irradiation temperature, $T_{\rm irr}$ \citep{Cowan2011-albedo, Komacek&Showman2016, Parmentier2018}: radiative timescale drops steeply with increasing temperature, following a $T^{-3}$ relation, so the hottest planets are expected to exhibit poor day-to-night transport. This general expectation is complicated by two additional phenomena -- high-temperature molecular dissociation and magnetic drag \citep{Bell2018,Menou2012}.

At the extreme temperatures of ultra-hot Jupiter atmospheres (T \textgreater 2300 K), H$_2$ can thermally dissociate on the dayside. On the nightside, it recombines, releasing latent heat that lowers the day-night temperature gradient \citep{Bell2018}. Indeed, the analysis by \cite{Mikal-Evans2022} of two full-orbit HST/WFC3 phase curves showed evidence for H$_2$O thermal dissociation and recombination matched by theoretical predictions \citep{Parmentier2018} and recent high-resolution studies \citep{Pelletier2025}. However, H$_2$O dissociation/recombination is not nearly as effective as H$_2$ for releasing latent heat, which is $\sim$ 100$\times$ more potent \citep{Bell2018}. Other factors, such as H$^-$ bound–free and free–free opacities shaping the continuum opacity \citep{Arcangeli2018} and enhanced magnetic drag slowing heat transport can further complicate the energy budget \citep{Beltz2022, Beltz2024}. Magnetic drag arises due to the extreme temperatures of ultra-hot Jupiters which leads to higher levels of thermal ionization of alkali metals that couple to the planet’s magnetic field. This interaction generates Lorentz forces that resist atmospheric motion, suppressing zonal winds and slowing day-to-night heat transport \citep{Perna2010,Menou2012,Tan2024, Kennedy2025}.

Previous phase curve observations of WASP-121\,b reveal significant discrepancies between observing instruments. For instance, estimates of Bond albedo and heat recirculation efficiency vary between different instruments due to differences in inferred dayside and nightside temperatures across their respective wavelength ranges \citep{Mikal-Evans2022, Morello2023, Dang2025}. Even within similar bandpasses, conflicting results emerge—Spitzer phase curve analyses at 3.6 and 4.5 $\mu$m \citep{Morello2023}, along with a subsequent 4.5 $\mu$m reanalyis \citep{Dang2025} suggest slight westward hot spot offsets, while a more recent reanalysis at both wavelengths finds an offset consistent with zero \citep{Davenport2025}. Meanwhile, phase offsets from NIRSpec NRS1 and NRS2 indicate a slight eastward shifts \citep{Mikal-Evans2023, Evans-Soma2025-NIRSpec-PhaseCurve}. However, uncertainties introduced by instrumental differences and correction methods make these conclusions tentative \citep{Changeat2024}.

A particular strength of NIRISS/SOSS is the ability to simultaneously probe reflected light and thermal emission \citep{Coulombe2025-LTT-NIRISS-PC, Morel2025}. Reflected stellar light, which can only be effectively probed at visible and near-infrared wavelengths, provides valuable insights into the presence and composition of aerosols in exoplanet atmospheres. Numerous studies of reflected light at optical wavelengths have revealed reflection on cooler, western portions of the planet where clouds are expected \citep{Demory2013, Parmentier2016, Coulombe2025-LTT-NIRISS-PC}. Moreover, an albedo paradox has been noted in hot Jupiters, where low geometric albedos are observed alongside higher Bond albedos \citep{Crossfield2015, Schwartz2015}. By capturing both reflected and emitted light, NIRISS/SOSS enables us to constrain the geometric albedo from reflected phase curves while simultaneously assessing the Bond albedo from thermal emission, thereby offering a comprehensive view of an exoplanet’s energy budget and atmospheric properties.

Here, we present JWST NIRISS/SOSS observations of the phase curve of WASP-121\,b, covering a spectroscopic wavelength range of 0.6–2.85\,µm. This range captures most of the planet's bolometric flux, enabling a more comprehensive energy budget analysis. Additionally, it overlaps with the wavelengths covered by TESS and Hubble and bridges the gap to the near-infrared coverage of NIRSpec/G395H and Spitzer/IRAC. By focusing on the NIRISS/SOSS bandpass, our analysis provides a critical comparison to previous observations, helping to reconcile discrepancies across different datasets.

In section \ref{sec:Methods} we describe our observations and the data reduction, the light-curve analysis and fitting, and energy balance models. In section \ref{sec:Results} we present the results of all our model fits and calculations. In section \ref{sec:Discussion} we disscuss the implications of our results for WASP-121\,b.

\vspace{10mm}
\section{Methods} \label{sec:Methods}

\subsection{Observations}

We obtained a full phase curve of WASP-121\,b, including two secondary eclipses and one primary transit, with JWST NIRISS/SOSS \citep{Albert2023,Doyon2023}. The observations were obtained as part of the NIRISS Exploration of the Atmospheric diversity of Transiting exoplanets (NEAT) Guaranteed Time Observation Program (GTO 1201; PI D. Lafrenière) on October 26th, 2023. Our time series observation lasted for 36.90 hours consisting of 6 groups per integration for a total of 3452 integrations. We used the SUBSTRIP256 subarray (256 × 2048 pixels) that covers the three spectral orders of the NIRISS instrument.

\subsection{Reduction}

We use two independent reduction pipelines to extract light-curves from the raw uncalibrated images: \texttt{exoTEDRF} \citep[formerly \texttt{supreme-SPOON};][]{exotedrf, Feinstein2023-WASP39b, Radica2023-AwesomeSOSS}  and \texttt{NAMELESS} \citep{Coulombe2023, Coulombe2025-LTT-NIRISS-PC}. The independent reductions produce white light-curves that are in good agreement with each other (Figure \ref{fig:exoTEDRF-NAMELESS}). We present the \texttt{exoTEDRF} reduction as our fiducial reduction.

\subsubsection{exoTEDRF}

For the \texttt{exoTEDRF} reduction, we used version 1.4.0 and followed the methodology outlined in \cite{radica_2024_muted} and \cite{radica_2025_promise}, as well as the official \texttt{exoTEDRF} tutorial\footnote{\url{https://exotedrf.readthedocs.io/en/latest/content/notebooks/tutorial_niriss-soss.html}}. The reduction process began with Stage 1, where we skipped the dark current subtraction step to prevent the introduction of additional noise.  

To correct for 1/$f$ noise, we applied the correction at the group level, prior to ramp fitting, by subtracting the background signal from each group. The background was removed using the STScI SOSS background model. After performing the 1/$f$ correction, the background was reintroduced to account for flat field and linearity effects before being removed once more for the final time.

We identified several contaminants that affected various steps in the reduction. Two bright order 0 contaminants were located at approximately $x \in [1000, 1150] $, $ y \in [125, 250] $, making background removal difficult to assess. An additional order 1 contaminant in the top-left corner impacted the 1/$f$ noise correction. Further contamination was observed near the spectral traces, including other order 0 contaminants and a possible order 1 contaminant. Contaminants at the top of the pixel frame interfered with both background subtraction and 1/$f$ noise correction. To ensure proper background removal, we used the default  background regions provided by \texttt{exoTEDRF}. Specifically we scaled two background regions from the STScI SOSS background model: ($x_1\in$ [350,550], $y_1\in$ [230,250]) and ($x_2\in$ [715,750], $y_2\in$ [235,250]), which were selected to avoid contamination. We examined row stacks before and after background extraction, evaluating pixel counts across 2048 spectral pixels. Despite contaminants increasing pixel counts at specific $ x $-positions, inspection of multiple rows along the $ y $-axis confirmed that background subtraction remained effective. We tested multiple background regions and found that was sufficient.  

To prevent the 1/$f$ noise correction from being biased by bad pixels, we carefully masked affected regions before computing the noise for each column. The pipeline's default masking automatically excluded the spectral traces and flagged bad pixels, which we further refined by manually identifying additional contaminated regions. Specifically, we masked pixels in the regions $ x\in[0,400],\, y\in[100,256]$ corresponding to the Order 1 contaminant and $ x\in[1050,1175],\, y\in[100,256]$ corresponding to the Order 0 contaminants. For the identified regions any pixels with flux values exceeding 2 (Order 1 contaminant region) or 3 (Order 0 contaminants region) times the local median were masked to reduce 1/$f$ noise correction impact.

As a final step, we corrected remaining bad pixels by interpolating outliers in both the spatial and temporal domains. For the spatial correction, we constructed a median stack of all integrations to identify pixels that deviated significantly from the stack across the entire time series. The temporal correction flagged pixels that appeared as outliers along the time axis. In both cases, pixels deviating by more than $10\sigma$ were replaced using the median of their surrounding pixels -- either within each integration (spatial) or across time (temporal)--resulting in a cleaner dataset.

Spectral trace centroids were determined using the edgetrigger algorithm \citep{Radica2022}. We then extracted the spectra using a simple box aperture, leveraging \texttt{exoTEDRF}'s built-in optimization to determine the ideal pixel width. A width of 29 pixels was found to minimize scatter, but we opted to extract with a width of 30 pixels. A box aperture extraction was chosen because self-contamination between spectral orders was found to be negligible during commissioning \citep{Darveau-Bernier2022, Radica2022}. However, we note the absence of a GR700XD/F277W exposure, which would typically aid in identifying contaminants hidden behind the spectral trace. The F277W filter, which transmits light redder than $ \sim 2.6 \,\mu$m, has proven effective for detecting such contaminants. Without this exposure, we lack a direct means to quantify dilution effects from hidden contaminants.

\subsection{Light-Curve Analysis}

Our model $F_{\rm model}(t)$ is a combination of an astrophysical model, $A(t)$, a systematics model, $S(t)$, and a Gaussian process model, $G(t)$, such that
\begin{equation}
F_{\rm model}(t) = A(t) + S(t) + G(t).
\end{equation}

\subsubsection{Astrophysical Model}

\noindent The astrophysical model is defined as the sum of stellar ($F_*$) and planetary flux ($F_p$):
\begin{equation}
A(t) = F_*(t) + F_p(t).
\end{equation}
\noindent Transits and eclipses are modeled using the \texttt{batman} Python package \citep{batman}. At shorter wavelengths (i.e., Order 2), the data are best described by a stellar variability model in which the flux varies sinusoidally with a period of 1.13 days, consistent with the stellar rotation period \citep{Delrez2016-discovery, Bourrier2020-stellar-rot}. At longer wavelengths (i.e., Order 1), the variability is negligible, and we assume a constant stellar flux. We therefore model the stellar flux as
\begin{equation}
F_*(t) =
\begin{cases} 
1 + A_{*}\sin\left(\frac{2\pi}{P_{*}}(t - t_{e}) - \phi_{*}\right) & \; \text{for Order 2} \\
1 & \; \text{for Order 1},
\end{cases}
\end{equation}
\noindent where $t_{e}$ is the time of the mid-eclipse, $P_{*}$ is the rotational period of the host star (assumed to be 1.13 days), $A_*$ is the amplitude of the model and $\phi_*$ is the stellar phase offset. 

The planetary flux, $F_p$ is modeled with respect to the orbital phase of the planet, $\theta$, by $\theta(t) = 2\pi(t - t_{e})/P_{\rm orb}$ ($P_{\rm orb}$ is the orbital period of the planet) such that 
\begin{equation} \label{eq:planetary_flux}
F_p(t) = F_{\rm day}\Phi(\theta(t)),
\end{equation} 
\noindent where $F_{\rm day}$ is the measured eclipse depth and $\Phi(\theta(t))$ describes phase variations. WASP-121\,b is expected to be significantly deformed by tidal forces \citep{Delrez2016-discovery,Daylan2021}. To accurately account for the planetary flux, we incorporate the variation in the planet's projected area as a function of orbital phase \citep[e.g.,][]{Bell2019, Kreidberg2018}. We follow \cite{Bell2019} and model our phase variations as: 
\begin{equation}
    \Phi(\theta(t)) = \Phi_{\rm p}(\theta)\Omega(\theta),
\end{equation}
\noindent where $ \Phi_{\rm p}$ describes the intrinsic brightness variations of the planet, and $\Omega(\theta(t))$ accounts for changes in the planet’s projected area due to its tidal deformation. For $ \Phi_{\rm p}$, we adopt a second-order sinusoid model following \cite{Cowan&Agol2008}:
\begin{align}
\Phi_{\rm p}(\theta)  &= 1 +   C_1 \left[ \cos\left( \theta \right) - 1 \right]+ D_1 \sin\left( \theta \right) \nonumber \\
&\; \;   + C_2 \left[ \cos\left(2 \theta \right) - 1 \right] + D_2 \sin\left(2 \theta \right)
\end{align}
\noindent To model the planet’s projected area, $\Omega(\theta)$, we assume a biaxial ellipsoid, where the polar and east--west axes are of equal length. This assumption is justified as rotational deformation is expected to be negligible compared to tidal deformation \citep{Leconte2011a}. Despite the relatively short stellar rotation period of WASP-121, the high obliquity of WASP-121\,b and the dominant tidal forces make rotational deformation unlikely to contribute significantly \citep{Delrez2016-discovery}. To model the projections of the biaxial ellipsoid, we adapt equations from previous studies \citep{Leconte2011b, Bell2019}:
\begin{align} 
\Omega(\theta) &= \left[ \sin^2(i) \left( \left( \frac{R_{\rm sub}}{R_p} \right)^2 \sin^2(\theta) + \cos^2(\theta) \right) \right. \notag \\
&\qquad \, \left. + \left( \frac{R_{\rm sub}}{R_p} \right)^2 \cos^2(i) \right]^{1/2},
\end{align}
\noindent where $i$ is the orbital inclination, $R_{\rm p}$ is the planetary radius along polar axis and $R_{\rm sub}$ is the planetary radius at the substellar point (line connecting the planet and the star).

\subsubsection{Systematics Model}

The systematics model corrects for potential trends in the data caused by instrument systematics or stellar variability. We make use of the optional functionality within \texttt{exoTEDRF} to apply a principal component analysis (PCA) to assess the stability of the SOSS trace during the course of the observation. More details of PCA analysis for SOSS can be found in \cite{Coulombe2023}. We used eigenvalues from the PCA analysis to detrend the data. In particular, we detrend PCA components 2--4 as the first component corresponds to the light-curve containing occultations and later components have minimal impact on the data. The second and third components likely correspond to drifts in the position of the trace, while the fourth component likely captures the characteristic beating pattern common to all SOSS observations, driven by the thermal cycling of the telescope's instruments.

The PCA analysis revealed the presence of a minor tilt event that occurs at the 3149th integration (see Figure \ref{fig:PCA-plot}), possibly due to a minor sudden movement of one of the telescope's primary mirror segments. A change in flux due to the tilt is not immediately noticeable in any of the light-curves. We nonetheless include a term in the systematics model to account for any potential consequences of the tilt event.

The systematics model is therefore formulated by a linear trend, a PCA eigenvalue model and a heaviside function to model tilt events such that
\begin{align}
    S(t) = a(t-t_{\rm mid}) + P_2\lambda_{\rm PCA_2} + \nonumber \\ P_3\lambda_{\rm PCA_3} + P_4\lambda_{\rm PCA_4} + jH(t-t_{\rm tilt}),
\end{align}
\noindent where $a$ represents the slope of the linear trend and $t_{\rm mid}$ corresponds to the middle of the observation (($t_{\rm min} + t_{\rm max}) / 2$). The eigenvalues of the $i$-th PCA component are represented by $\lambda_{\rm PCA_i}$ and $P_i$ is the amplitude used for detrending. To avoid reintroducing any long-term effects from other systematic models each PCA eigenvalue is smoothed by subtracting a rolling median with a window size of 50 \citep{Coulombe2025-LTT-NIRISS-PC}. We modeled the tilt event in the light-curve fitting with a Heaviside function at the time of the tilt event ($t_{\rm tilt}$) where $j$ is amplitude of the jump. 

\subsubsection{Gaussian Process} \label{sec:GaussianProcess}

We implement a Gaussian process (GP) model to marginalize our inferences over leftover nonwhite noise. We use the open-source Python package \texttt{celerite} \citep{celerite}, and consider the SHOTerm Kernel which represents a stochastically driven, damped harmonic oscillator. This kernel models the power spectral density, $S(\omega)$, using three input parameters: the signal amplitude $S_0$, characteristic frequency $\omega_0$, and quality factor $Q$:

\begin{equation}
S(\omega) = \sqrt{\frac{2}{\pi}} \frac{S_0 \, \omega_0^4}{(\omega^2 - \omega_0^2)^2 + \omega_0^2 \, \omega^2 / Q^2}.
\end{equation}

\noindent The quality factor is set to \( Q = 1/\sqrt{2} \), a commonly used value for modeling stellar granulation. We use the scaling relations of \cite{Gilliland2011} (equations 5 and 6 of that work) and find an expected granulation signal timescale of 6 minutes and an amplitude of 96 ppm, a level we should be sensitive to. Following \citet{radica_2024_muted} and \cite{Coulombe2025-LTT-NIRISS-PC}, we adopt the parameterization of \cite{Pereira2019-Granulation} and reframe the GP parameters in terms of amplitude, \( A_{\rm GP} = \sqrt{\sqrt{2} S_0 \omega_0} \), and timescale, \( \tau_{\rm GP} = 2\pi/\omega_0 \). These are then converted back to \( S_0 \) and \( \omega_0 \) before being passed to \texttt{celerite}.

\subsubsection{Light-Curve Fitting}

We fit our light-curves using the \texttt{emcee} Python package, which implements a Monte Carlo ensemble sampler with affine-invariant Markov chains \citep{emcee}. We construct the white light-curves by summing the flux across all 2048 pixels [0.85--2.85 $\mu$m] for Order 1 and from [0.6--0.85 $\mu$m] for Order 2. We then normalize both white light-curves to the first eclipse and are then analyzed separately, with the orbital eccentricity fixed to zero \citep{Delrez2016-discovery,Bourrier2020}.

For each order, we fit a total of 21–23 parameters. The orbital parameters include the orbital period ($P$), the semi-major axis relative to the stellar radius ($a/R_\ast$), the mid-transit time ($T_0$), and the orbital inclination ($i$). The astrophysical parameters consist of the planet-to-star radius ratio ($R_{\rm p}/R_\ast$), the substellar planet radius to stellar radius ratio ($R_{\rm sub}/R_\ast$), the quadratic limb-darkening coefficients ($q_1, q_2$) \citep{Kipping2013}, and the eclipse depth ($F_{\rm day}$). To account for phase variations, we fit four sinusoidal parameters ($C_1, C_2, D_1, D_2$). To avoid unphysical solutions, we impose a prior on the planetary flux (equation \ref{eq:planetary_flux}). Specifically, for any modeled point that yields negative flux, we assign a half-Gaussian prior centered at zero, with the width $\sigma$ equal to the measurement uncertainty of that point. This construction allows the sampler to explore negative flux values, but increasingly penalizes them in proportion to how many sigma below zero they fall, thereby favoring physically plausible (positive) fluxes without imposing a hard boundary. Additionally, we include five systematics parameters: a linear slope ($a$), three principal component analysis amplitudes ($P_2, P_3, P_4$), and the tilt-event jump amplitude ($j$). We also fit for the Gaussian process amplitude ($A_{\rm GP}$) and timescale ($\tau_{\rm GP}$), along with a scatter parameter ($\sigma$) that is added in quadrature to the measurement uncertainties.

A Bayesian information criterion (BIC) analysis strongly favors the inclusion of the stellar variability model in the Order 2 white light-curve ($\Delta \text{BIC} \gtrsim 33$), while there is no preference for its inclusion in the Order 1 white light-curve ($\Delta \text{BIC} \lesssim 2$). A sinusoidal function best describes this variability, leading us to introduce two additional parameters: the stellar variability amplitude ($A_\ast$) and phase offset ($\phi_\ast$). As a result, the total number of fitted parameters is 21 for the Order 1 white light-curve and 23 for Order 2.

For each fit, we initialize the sampler with four walkers per parameter and run for 50,000 steps. We first conduct an initial fit with 12,500 steps, corresponding to 25\% of the total run, and use the maximum-probability parameters from this preliminary fit as the starting positions for the final 50,000-step run. To ensure convergence, we discard the first 30,000 steps, or 60\% of the total run, as burn-in, retaining only the posteriors from fully converged walkers. We check convergence through visual inspection of trace plots and corner plots, as well as by computing the integrated autocorrelation time \citep{emcee}. For both orders, the retained number of steps per walker exceeded 50 times the autocorrelation length.

We fit our spectral light-curves by binning the data by pixel with 5 pixels per bin and again normalize the light-curves to the first eclipse. For Order 2, we exclude wavelengths longer than $0.85 \, \mu$m to prevent overlap with Order 1. This results in 113 spectral light-curves for Order 2 and 409 for Order 1.

We fix the four orbital parameters to the median values obtained from the posteriors of white light-curve analysis: $t_0 =60244.520345$ MJD for Order 1, $t_0 =60244.520412$ MJD for Order 2, $P_{\rm orb} = 1.274925$ days, $a/R_* = 3.8002$ and $i= 88.345^\circ$. Systematics parameters are assigned Gaussian priors centered on their best-fit values from the white light-curve fit, with standard deviations set to the corresponding measurement uncertainties. Rather than fitting a separate GP model at each wavelength, we scale the mean GP model from the white light-curve fit by introducing a scaling factor, $k_{\rm GP}(\lambda)$ \citep[e.g.,][]{radica_2024_muted,Coulombe2025-LTT-NIRISS-PC}. Additionally, we fix the projected area of the planet, determined from $R_{\rm sub}/R_\ast$, based on the fit of Order 1 white light-curve ($R_{\rm sub}/R_\mathrm{p} = 1.157 \pm 0.023$). This value is consistent within 1$\sigma$ of the result from \cite{Delrez2016-discovery} ($R_{\rm sub}/R_\mathrm{p} = 1.124 \pm 0.026$). For the stellar variability model in Order 2, we fix the phase offset to the best-fit value from the white light-curve fit and constrain the amplitude using a Gaussian prior informed by the white light-curve results.

This leaves eight astrophysical parameters as free parameters at each wavelength: the planet-to-star radius ratio ($R_{\rm p}/R_\ast$), the quadratic limb-darkening coefficients ($q_1, q_2$), the eclipse depth ($F_{\rm day}$), and the sinusoidal phase variation parameters ($C_1, C_2, D_1, D_2$) which are assigned wide uniform priors to allow for flexibility in the fit.

We adopt the same fitting strategy as for the white light-curve. The first spectroscopic bin is fit using 50,000 iterations. For subsequent wavelength bins, the best-fit parameters from the previous bin are used as initial conditions, allowing the sampler to converge more efficiently. These subsequent bins are ran for 12,500 iterations, as they begin closer to their optimal solutions. As in the white light-curve fit, each spectroscopic bin is first run for an initial burn-in phase consisting of 25\% of the total iterations, after which the fit is restarted at its maximum-likelihood parameters for the full iteration count. To ensure robust posterior sampling, we discard the first 60\% of steps as burn-in before analyzing the final parameter distributions.

In both our fits for the white light-curve and spectroscopic light-curve fits we allow the astrophysical model to fit for negative planetary flux but disfavor such solutions by applying a half-Gaussian prior as described above.

\begin{figure*}[]
        \includegraphics[width=0.995\textwidth]{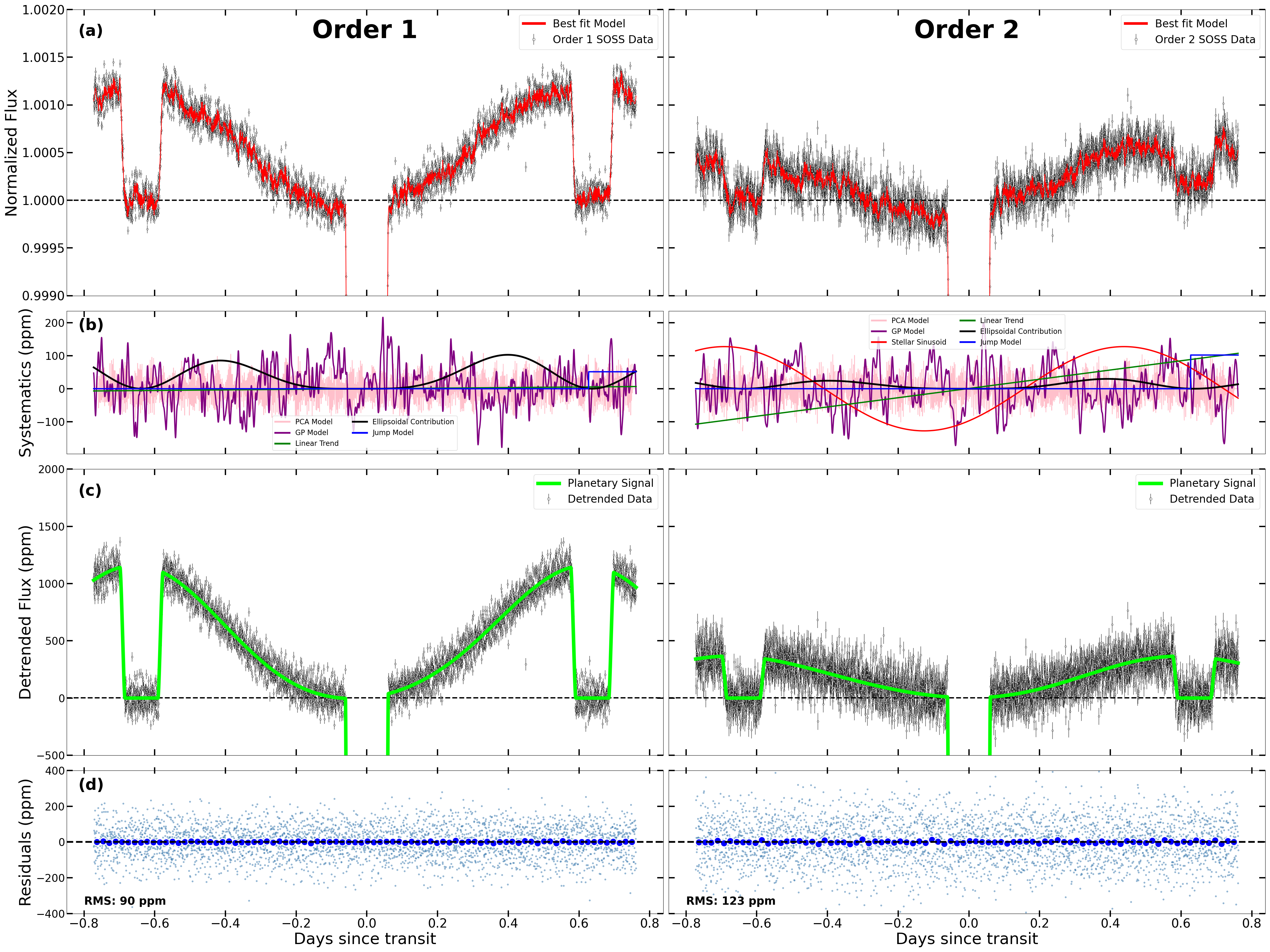}
    
    \caption{\textbf{Left:} \emph{(a)} Order 1 white light-curve fit to the flux that is normalized to the first eclipse. The red line is the best-fit full model. \emph{(b)} Components of the systematics model in ppm, including the linear trend in green; the PCA model in pink; the tilt jump model in blue; the GP model in purple; the ellipsoidal variation contribution in black; and the stellar sinusoid model in red (Order 2 only). \emph{(c)} Detrended light-curve created after removing all contributions from (b). The remaining planetary signal is shown in green. \emph{(d)}    Residuals of the best-fit model to the data. Darker blue points are residuals binned for visual purposes. The rms in ppm is printed in the bottom left. \textbf{Right:} same as the left plot except for Order 2.}
    \label{fig:WLC-fits}
\end{figure*}

\begin{table*}[htb]
\caption{Retrieved Median Values from MCMC Fit on Order 1 and Order 2}
\label{tab:mcmc_results_combined}
\centering 
\resizebox{\linewidth}{!}{
\begin{tabular}{lcccc}
\toprule
Parameter & Symbol [Unit] & Order 1 & Order 2 & Prior \\
\midrule
Time of Inferior Conjunction & \(t_0\) [MJD] & \(60244.520345 \,\pm 0.000032\) & \(60244.520412\,^{+0.000039}_{-0.000040}\) & \(\mathcal{U}(60244.51,\,60244.53)\) \\
Orbital Period & \(P_{\rm orb}\) [day] & \(1.274925\,\pm 1.5\times10^{-7}\) & \(1.274925\,\pm 1.5\times10^{-7}\) & \(\mathcal{N}(1.27492504,\, 1.5\times10^{-7})^{\rm a}\) \\
Planetary Radius & \(R_{\rm p}/R_*\) & \(0.1222\,\pm 0.00014\) & \(0.12293\,\pm 0.0002\) & \(\mathcal{U}(0,\,1)\) \\
Semi-major Axis & \(a/R_*\)  & \(3.8002\,\pm 0.005\) & \(3.8023\,^{+0.0061}_{-0.0063}\) & \(\mathcal{N}(3.754,\, 0.028)^{\rm b}\) \\
Inclination & \(i\) [deg] & \(88.345\,^{+0.158}_{-0.157}\) & \(88.39\,\pm 0.16\) & \(\mathcal{N}(88.49,\, 0.16)^{\rm a}\) \\
Eclipse Depth & \(F_{\rm Day}\) [ppm] & \(1155.99\,^{+14.30}_{-14.04}\) & \(363.13\,^{+20.79}_{-21.64}\) & \(\mathcal{U}(-500,\,1\times10^6)\) \\
Substellar Radius & \(R_{\rm sub}/R_*\) & \(0.1414\,\pm 0.0028\) & \(0.139\,\pm 0.0032\) & \(\mathcal{U}(0,\, 1)^{\rm b, \, \dagger}\) \\
Linear Slope  & \(a\) [ppm/day] & \(8.68\,^{+15.51}_{-15.63}\) & \(140.18\,^{+25.70}_{-25.15}\) & \(\mathcal{U}(-1000,\,1000)\) \\
Jump Amplitude & \(j\) [ppm] & \(51.55\,^{+22.26}_{-21.65}\) & \(102.27\,\pm {28.06}\) & \(\mathcal{U}(-1000,\,1000)\) \\
GP Period & \(\log_{10}{\tau_{\rm GP}}\) [$\log_{10}$(min)] & \(1.478\,\pm {0.06}\) & \(1.549\,\pm 0.069\) & \(\mathcal{U}(\log_{10}(1),\,\log_{10}(100))\) \\
GP Amplitude & \(\log_{10}A_{\rm GP}\) [$\log_{10}$(ppm)] & \(1.917\,\pm 0.027\) & \(1.976\,\pm 0.03\) & \(\mathcal{U}(\log_{10}(10),\,\log_{10}(500))\) \\
Scatter & \(\sigma\) [ppm] & \(88.8\,^{+1.46}_{-1.43}\) & \(106.7\,^{+2.06}_{-2.07}\) & \(\mathcal{U}(0,\,10^9)\) \\
Stellar Offset & \(\phi_*\) [rad] & -- & \(3.61\,^{+0.26}_{-0.19}\) & \(\mathcal{U}(-\pi,\,2\pi)\) \\
Stellar Amplitude & \(A_*\) [ppm] & -- & \(128\,^{+26.5}_{-21.3}\) & \(\mathcal{U}(0,\,1000)\) \\
\bottomrule
\multicolumn{5}{r}{$^{\rm a}$\citet{Bourrier2020}, $^{\rm b}$\citet{Delrez2016-discovery}} \\
\multicolumn{5}{r}{$^{\rm \dagger}$Additional prior on $R_{\rm sub}/R_{\rm p}$ of $\mathcal{N}(1.124,\, 0.026)$ from \citet{Delrez2016-discovery}}
\end{tabular}
}
\end{table*}

\subsubsection{Differences between exoTEDRF and NAMELESS light-curve fits} \label{sec:reduction-fit-differences}

Here, we have described the reduction and light-curve fitting process used in our analysis with \texttt{exoTEDRF}. As mentioned earlier, an independent reduction and light-curve analysis using \texttt{NAMELESS} is detailed in \cite{Allart2025} and \cite{Pelletier2025}, with a brief summary provided in Appendix \ref{app:nameless}. However, we provide a brief overview of key differences in the fitting approaches for comparison later in this work.

There are three primary methodological differences between the two analyses:
\begin{enumerate}

\item The \texttt{exoTEDRF} analysis includes ellipsoidal variation in the light-curve model, whereas \texttt{NAMELESS} does not.

\item Both fits allow for negative planetary flux however, the \texttt{exoTEDRF} fit penalizes negative flux values with a half-gaussian prior while the \texttt{NAMELESS} fit has a completely uninformative uniform prior.

\item For Order 2, the \texttt{exoTEDRF} model incorporates a stellar variability sinusoid, unlike the \texttt{NAMELESS} model, which does not.
\end{enumerate}

These differences have several implications. While the contribution of ellipsoidal variation to the total flux is found to be small, neglecting it can lead to a slight overestimation of flux at orbital quadrature. The handling of negative flux presents a more complex challenge. On the one hand, negative planetary flux is physically unrealistic, requiring systematic trends to compensate for observed dips. On the other hand, strictly enforcing positive planetary flux risks misrepresenting the true shape of the phase curve. By applying a half-gaussian penalty on negative planetary flux, we aim to bridge the gap between the two approaches. The effects of these fundamental fitting differences are seen most prominently when estimating the nightside flux (see Figures \ref{fig:Day-v-nightside}). Further, as a negative flux cannot be converted into a brightness temperature, the nightside temperature will have a larger uncertainty (Figure \ref{fig:Nightside-Brightness-Temp}). 

Despite these methodological differences, comparing these two independent analyses provides valuable insight into how different modeling approaches influence the interpretation of the data. Each method carries its own biases, and examining their deviations allows us to better assess the robustness of our results.

\subsection{Slice Model} \label{sec:Slice-Model}

The NIRISS/SOSS wavelength coverage extends into both the optical and near-infrared, allowing us to estimate the planet's reflected light. If we assume the planet has a gray albedo and emits as a blackbody, a sufficiently high geometric albedo ($\sim$ 0.2) would result in reflected light dominating at lower wavelengths, likely within Order 2 (see Fig \ref{fig:Estimated-reflected-light}). To distinguish reflected light from thermal emission, we decompose the detrended, phase-dependent planetary flux into these two components:
\begin{equation}
   F_p(t) =  F_r(t) + F_e(t).
\end{equation}
We model these components using a longitudinal slice approach, following methods applied in previous studies \citep{Cowan&Agol2008,Coulombe2025-LTT-NIRISS-PC}. The planet is divided into $N$ longitudinal slices, where each slice is characterized by a thermal emission component, $T_i$, and an apparent albedo, $A_i$. The total planetary flux is then expressed as:
\begin{equation} \label{eq:slice_model}
   F_p(t,\lambda) =  \sum^N_{i=1} \bigg[A_i \,K_{{\rm reflect} ,\, i}(t) + T_i \,K_{{\rm thermal} ,\, i}(t) \bigg],
\end{equation}
where $K_{\rm reflect}$ and $K_{\rm thermal}$ are convolution kernels for the reflected and thermal components, defined by integrals over the solid angle of the planetary longitude $\phi$ and latitude $\theta$:
\begin{equation} \label{eq:reflect_kernel}
   K_{{\rm reflect} ,\, i} = F_*(\lambda) \left(\frac{R_*}{a} \right)^2 \frac{1}{\pi} \int^{\phi_{i+1}}_{\phi_i} V(\theta,\phi, t) I(\theta, \phi, t)d\Omega , 
\end{equation}
\begin{equation} \label{eq:emission_kernel}
   K_{{\rm thermal} ,\, i} = \frac{1}{\pi} \int^{\phi_{i+1}}_{\phi_i} V(\theta,\phi, t)d\Omega.
\end{equation}
\noindent Here, $V(\theta,\phi, t)$ and $I(\theta,\phi, t)$ represent the normalized visibility and illumination of a given planetary location at time $t$ \citep{Cowan2013}. 

The visibility, $V$, quantifies which regions of the planet are visible to the observer: 
\begin{equation} \label{eq: vis}
    V(\theta, \phi, t) = \max[\sin\theta\sin\theta_{o}\cos(\phi-\phi_{o})+\cos\theta\cos\theta_{o},\, 0],
\end{equation} 
where $\theta_{o}$ and $\phi_{o}$ are the sub-observer latitude and longitude respectively. The illumination, $I$, describes the illumination of different regions by the host star:
\begin{equation} \label{eq: illumin}
    I(\theta, \phi, t) = \max[\sin\theta\sin\theta_{s}\cos(\phi-\phi_{s})+\cos\theta\cos\theta_{s}, \,0].
\end{equation}
We define the planetary longitude $\phi \in [0,2\pi]$ and latitude $\theta \in [0,\pi]$ such that the substellar position is at $(\phi_s = 0, \,  \theta_s = \frac{\pi}{2})$. The longitudes seen by the observer change with time, $\phi_o = -2\pi (t-t_{\rm e})/P_{\rm orb}$, while the observed latitude is equal to the inclination ($i \sim 88.3^\circ$) which we approximate as edge-on, $\theta_s =  \frac{\pi}{2}$.

We analytically solve for both kernels and use these solutions to precompute the reflected and thermal component weights. This allows equation \ref{eq:slice_model} to be reduced to a simple matrix multiplication, where the computed flux has a shape of $N_{\rm time}$. The albedo ($A_i$) and thermal emission ($T_i$) parameters have a shape of $N_{\rm slice}$, while the kernel weights are structured as ($N_{\rm slice}$, $N_{\rm time}$). This precomputation significantly accelerates our calculations, which is essential since the longitudinal slices are at least partially degenerate with one another. Consequently, the fits require more steps and walkers to ensure proper convergence.  

To address this, we follow a similar approach to our sinusoidal fits using \texttt{emcee}, but we increase the total number of steps to 100,000 and use 100 walkers. Naïvely, the fit would include $2N_{\rm slice} + 1$ parameters: $N_{\rm slice}$ for the albedo values, $N_{\rm slice}$ for the emission parameters, and one additional scatter parameter, $\sigma$. However, since nightside slices do not contribute  to the reflected light component, we exclude these albedo values from the fit. In any case, our choice of 100 walkers ensures a sufficient number of walkers per free parameter. Following \cite{Coulombe2025-LTT-NIRISS-PC} we set an upper prior limit of $3/2$ on all albedo slices as a fully Lambertian sphere ($A_{i} = 1$) corresponds to a geometric albedo of $A_\mathrm{g} = 2/3$. For thermal emission we impose a uniform prior between 0 and 500 ppm for each slice.

We choose to fit our detrended light curves considering four, six and eight longitudinal slices ($N_{\rm slice}=4,6,8$). However, we show the results of the simplest four slice model. As in our previous fits, we conduct an initial run with 25,000 steps (25\% of the total run) and use the maximum-probability parameters from this preliminary fit as the starting positions for the final 75,000-step run. We then discard the first 60\% of the final run as burn-in.

\subsection{Planetary Effective Temperature} \label{sec:T-eff-creation}

\begin{figure}[h]
    \centering
    \includegraphics[width=0.495\textwidth]{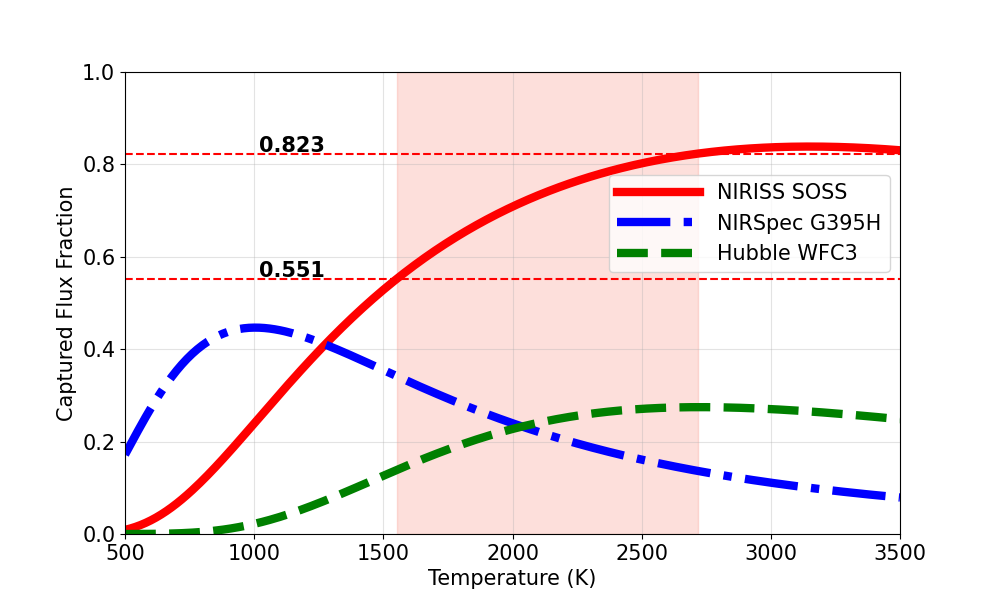}
    
	\caption{Estimated captured flux of the planet assuming the planet radiates as a blackbody. The captured flux is calculated as the ratio of the integrated blackbody emission within the instrument's bandpass to the total emission over all wavelengths, i.e., $\gamma =\int_{\lambda_{\rm min}}^{\lambda_{\rm max}} B(\lambda, T)\, d\lambda / \int_{0}^{\infty} B(\lambda, T)\, d\lambda$. The captured flux fraction is shown for NIRISS SOSS [0.6–2.85 $\mu$m] (red line); Hubble WFC3 [1.12–1.64 $\mu$m] (dashed green line); NIRSpec G395H [2.7–5.15 $\mu$m] (dash-dotted blue line). The red-shaded region shows the temperature range on WASP-121\,b based on our $T_{\rm eff}$ estimates. Red dashed lines indicate the boundaries of the planet's temperature range within the NIRISS SOSS captured flux fraction. From this we estimate that these observations capture between 55\% and 82\% of the planet's bolometric flux, depending on orbital phase. Using the minimum temperature from the \texttt{NAMELESS} fit, this estimate decreases to 50\%. In either case, the wavelength coverage of NIRISS exceeds that of any other instrument.}
	\label{fig:Captured-Flux-Fraction}
\end{figure}

Phase curves are the only way to probe thermal emission from the day- and nightside of an exoplanet and hence determine its global energy budget \citep{Phasecurve-book}. The wavelength range of NIRISS/SOSS covers a large portion of the emitted flux of WASP-121\,b ($\sim$ 50--83\%; see Figure \ref{fig:Captured-Flux-Fraction}), enabling a precise and robust constraint of the planet's energy budget.

We convert the fitted $F_p / F_*$ emission spectra to brightness temperature by wavelength,
\begin{align}
    T_{\rm bright} = \frac{h c}{k \lambda} \cdot \left[ {\ln\left(\frac{2 h c^2}{\lambda^5 B_{\lambda , {\text{planet}}}} + 1\right)} \right]^{-1},
\end{align}
\noindent where the planet's thermal emission is 
\begin{align}
B_{\lambda ,\, \text{planet}} = \frac{F_{p}/F_{*}}{(R_p/R_*)^2} \cdot B_{\lambda ,\,  \text{star}} \, .
\end{align}

There are many ways of converting brightness temperatures to effective temperature, including the Error-Weighted Mean (EWM),  Power-Weighted mean (PWM) and with a Gaussian process \citep{Schwartz2015, Pass2019-TeffGP}. In this work, we elect to compute our effective temperature estimates with a novel method that is essentially a combination of the PWM and EWM. We create the effective temperature by using a simple Monte Carlo process. First, we perturb our $F_p / F_s$ emission spectra at each point in the orbit by a Gaussian based on the measurement uncertainty. Our new emission spectrum is then used to create an estimate of the brightness temperature spectrum. This process is repeated at each orbital phase. We then estimate the effective temperature, $T_{\rm eff}$ for a given orbital phase as
\begin{align}
   T_{\rm eff} = \frac{\sum_{i=1} ^{N}w_i \, T_{{\rm bright}, \,i} }{\sum_{i=1} ^{N}w_i},
\end{align}
\noindent where $w_i$ is the weight for the $i$-th wavelength given by the fraction of the planet's bolometric flux that falls within that wavelength bin scaled by the inverse variance of the measurement,  
\begin{align}
   w_i = \frac{\int^{\lambda_{ i+1}}_{\lambda_{ i}} B(\lambda_i,\, T_{{\rm est}}) \, d\lambda}{\int^{\infty}_{0} B(\lambda_i,\, T_{{\rm est}}) \, d\lambda} \cdot \frac{1}{\sigma_{i}^2}, 
\end{align}
\noindent with $T_{{\rm est}}$ representing an estimated effective temperature at the orbital phase of interest. When computing $T_{{\rm est}}$, we use the EWM of the Order 1 wavelengths to ensure that only thermal emission is considered.

We run the Monte Carlo simulations $1000$ times, propagating all relevant astrophysical uncertainties. Specifically, we propagate the uncertainties of $a/R_*$ and $R_{\rm p}/R_*$ as determined by the Order 1 white light-curve fit ($a/R_* = 3.8002\,\pm {0.0005}$, $R_{\rm p}/R_* = 0.1222\,\pm{0.00014}$). Additionally, we account for the uncertainty in the stellar effective temperature, adopting the value reported in \cite{Sing2024-NIRSpec} ($T_* = 6628 \pm 66 $ K). We record the median and 1$\sigma$ interval of the effective temperature at all orbital phases.

Finally, we inflate the uncertainties of the effective temperatures by the total captured flux fraction of a blackbody at the effective temperature across the NIRISS/SOSS bandpass (see Figure \ref{fig:Captured-Flux-Fraction}). If we let $\gamma = F/F_{\rm bol}$ be the captured flux fraction, our new, inflated uncertainty on the effective temperature becomes $\sigma^\prime = \sigma/\gamma$. If the entirety of the planetary SED is captured, the uncertainties are unchanged, whereas if we only capture half of the bolometric flux, then the uncertainties are doubled.

\section{Results} \label{sec:Results}

\subsection{Bond Albedo and Heat Recirculation Efficiency}

\begin{figure}[htb]
    \centering
    \includegraphics[width=0.495\textwidth]{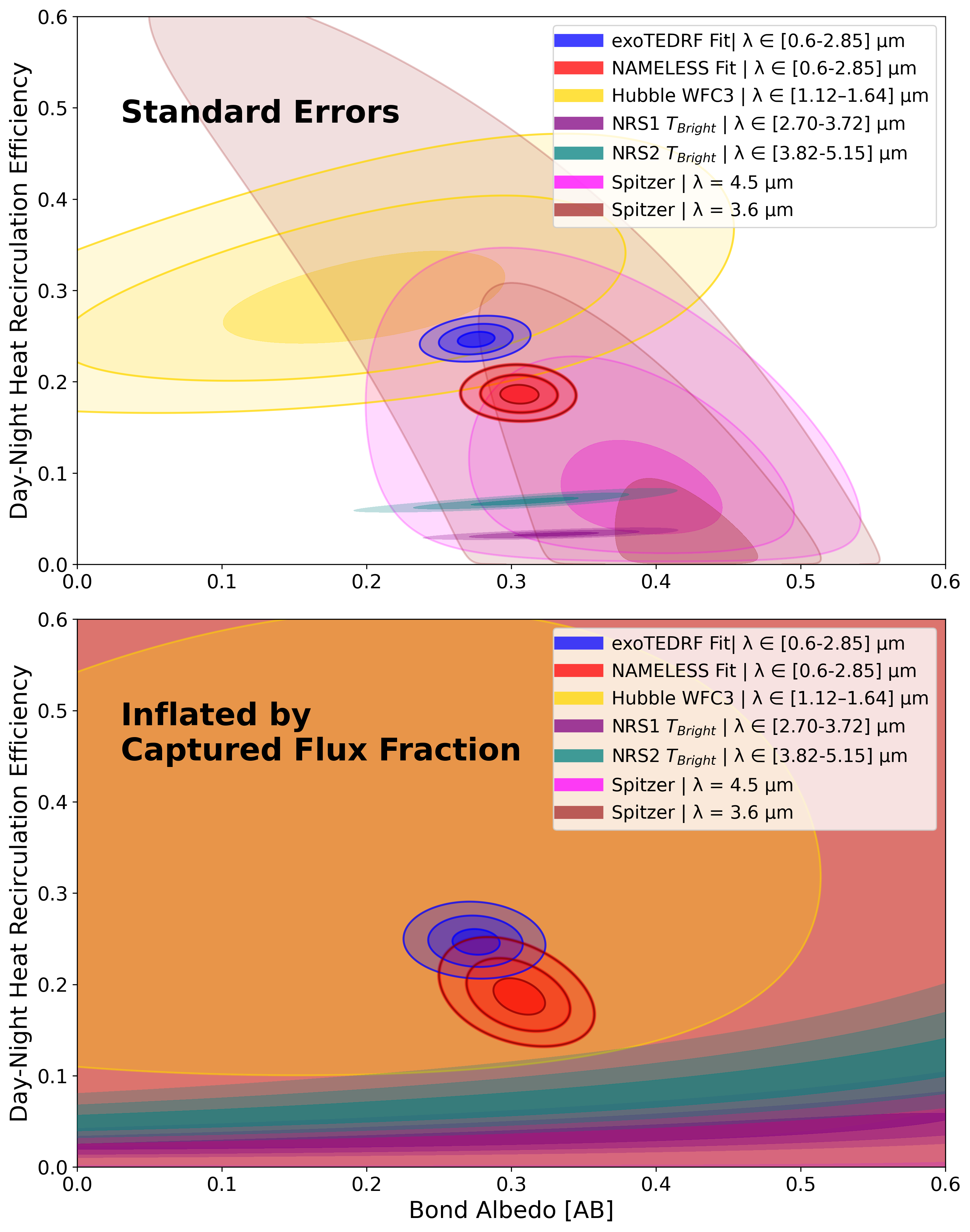}
    
	\caption{Inferred Bond albedo (A$_\mathrm{B}$) and heat recirculation efficiency ($\epsilon$) of WASP-121\,b derived from dayside and nightside emission measurements using irradiation temperature $T_{0} = 3398$\,K.  The blue contours represent 1$\sigma$, 2$\sigma$ and 3$\sigma$ confidence intervals from the fit to the \texttt{exoTEDRF} reduction, while the red contours correspond to the \texttt{NAMELESS} reduction. The gold contours show the results from the Hubble WFC3 phase curve presented by \cite{Mikal-Evans2022}, using the brightness temperatures reported in \cite{Morello2023}. The purple and teal contours correspond to brightness temperatures from NIRSpec NRS1 and NRS2, respectively \citep{Mikal-Evans2023}. For the Spitzer 3.6 $\mu$m phase curve (brown contours), we use brightness temperatures from \cite{Morello2023}; while the magenta contours represent results from the comprehensive analysis of the 4.5~$\mu$m phase curve by \cite{Dang2025}. \emph{Top:} Inferred contours when using standard uncertainties of day- and nightside. Only NIRISS SOSS measurements fully propogate astrophysical uncertainties. \emph{Bottom}: Inferred contours when uncertainties are inflated by 1/$\gamma$, i.e. the captured flux fraction for a given instrument at the reported temperature. When accounting for captured flux, only NIRISS provides constraints on both the Bond albedo and heat recirculation efficiency. Our phase measurements show a Bond albedo of 20--35\% and relatively poor day-to-night heat transport. 
}
	\label{fig:albedo-recirculation}
\end{figure}

Using the $T_{\rm eff}$ values for the dayside ($T_{\rm day} = 2717 \pm 17$K) and nightside ($T_{\rm night} = 1562^{+18}_{-19}$K) hemispheres (where we have inflated the uncertainties by $\gamma$), we infer solutions of the Bond albedo  $A_\mathrm{B}$ and the day-night heat recirculation efficiency $\epsilon$ following \cite{Cowan2011-albedo}: 
\begin{align} \label{equation:albedo-recirc}
T_{\rm day} = T_0 (1 - A_\mathrm{B})^{1/4} \left(\frac{2}{3} - \frac{5}{12}\epsilon\right)^{1/4}, \\ \notag T_{\rm night} = T_0 (1 - A_\mathrm{B})^{1/4} \left(\frac{\epsilon}{4}\right)^{1/4},
\end{align}
where $T_0$ is the irradiation temperature and $A_\mathrm{B}$ and $\epsilon$ range from 0 to 1. We recorded the irradiation temperature during the Monte Carlo simulation described in Section \ref{sec:T-eff-creation} and find $T_0 = T_{\ast} \sqrt{R_\ast/a} = 3398\,^{+35}_{-34}$ K.

Following \cite{Schwartz2015}, we generate a 2000 $\times$ 2000 grid of day- and nightside temperatures as a function of the Bond albedo and heat recirculation efficicency. We then create a $\chi^2$ surface from equation \ref{equation:albedo-recirc} by comparing these models to our measured dayside and nightside effective temperatures. In order to properly account for the uncertainty in irradiation temperature we recorded $\frac{T_{\rm day}}{T_{0}}$ and $\frac{T_{\rm night}}{T_{0}}$ during each iteration of the Monte Carlo simulation from section \ref{sec:T-eff-creation}. We then find the $1\sigma$, $2\sigma$, and $3\sigma$ confidence intervals defined when $\Delta\chi^2 < [1,4,9]$ respectively. For our fit, we infer values of  $A_\mathrm{B} = 0.277\, \pm0.016$ and $\epsilon = 0.246\, \pm 0.014$ from temperatures with uncertainties inflated by $1/\gamma$. 

We compare our results of $A_\mathrm{B}$ and $\epsilon$ with past phase curve observations of the planet. Previous reported bond albedos and heat recirculation efficiencies were estimated using different irradiation temperatures. To homogenize our comparisons, we use the reported dayside and nightside temperatures from each study and use our calculated irradiation temperature. However, we do not propagate uncertainty in irradiation temperature for other instruments. A fully rigorous error propagation would require reprocessing the observed emission spectra with a consistent stellar effective temperature and irradiation temperature at each Monte Carlo iteration, as done in Section \ref{sec:T-eff-creation}. The results of our energy budget with that of previous studies is shown in Figure \ref{fig:albedo-recirculation}.

\subsection{Phase Offsets and Relative Amplitude}

\begin{figure}[htb]
    \centering
    \includegraphics[width=0.495\textwidth]{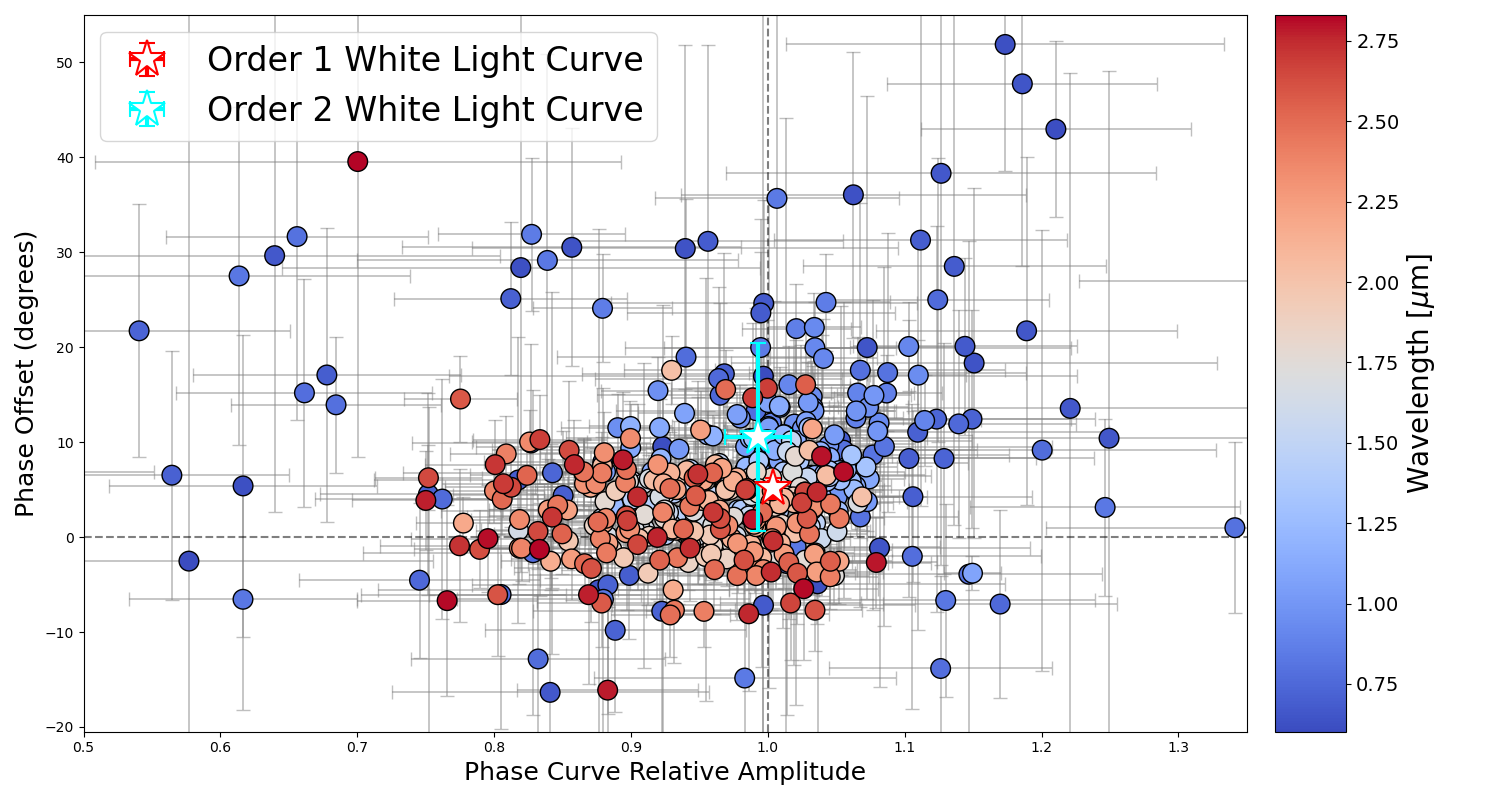}
    
	\caption{Phase offset versus normalized phase amplitude from the spectroscopic fits of WASP-121\,b. The normalized phase amplitude is given by $(F_{\rm max} - F_{\rm min})/F_{\rm max}$. The phase offset is defined such that 0$^\circ$ corresponds to mid-eclipse with positive offset values indicating an eastward shift and negative a westward. Symbols are colored by wavelength with blue indicating the shortest wavelengths and red the longest. Order 2 values are binned with three points per bin for visual clarity. Relative amplitudes above 1 correspond to Order 2 wavelengths with a negative minimum flux. The values from the white light-curve fits are also shown as stars with colored cyan edges for Order 2 and red for Order 1. We robustly detect an eastward offset of 5.1 $\pm$ 1.4$^\circ$ in Order 1 and an insignificant offset of 10.5 $\pm$ 9.9$^\circ$ in Order 2.}
	\label{fig:dobbs-dixon}
\end{figure}

We use our light-curve fits to estimate the phase offset and the phase curve relative amplitude. The phase offset is defined as the orbital phase of the brightest location of the light-curve relative to the substellar point. To first order, it is the longitude of the hot spot \citep{Schwartz2017}. Positive phase offsets correspond to an eastward shift (i.e., maximum flux occurs before mid-eclipse), while negative offsets indicate a westward location. Meanwhile, the phase curve relative amplitude is the normalized amplitude between the maximum flux and the minimum flux (i.e., $(F_{\rm max} - F_{\rm min})/F_{\rm max}$). This metric serves as a proxy for the day–night temperature contrast; for example, a relative amplitude of 1 would indicate that one hemisphere emits no detectable flux. Such measurements offer critical insight into the energy balance and heat redistribution in the planetary atmosphere \citep{Cowan2012, Schwartz2015}.

We compare the relative phase amplitude and phase offsets at different wavelengths in Figure \ref{fig:dobbs-dixon}. \citet{Dobbs-Dixon&Cowan2017} used a similar figure to show that hot Jupiters observed with Spitzer exhibit larger relative amplitudes and phase offsets at 3.6 $\mu$m compared to 4.5 $\mu$m. In an idealized bolometric light-curve, we naively might expect phase offset and amplitude to have a one-to-one relationship, where planets with larger phase amplitudes (stronger day-night temperature contrasts) should exhibit smaller phase offsets \citep{Schwartz2017}. However, more recent work shows that hot Jupiters are diverse and that the phase offsets and amplitudes are not necessarily correlated and can be shaped by different components of the circulation \citep{Roth2024}.

We find phase offsets both shifted eastward in the white light-curves as 5.1 $\pm$ 1.4$^\circ$ for Order 1 and 10.5 $\pm$ 9.9$^\circ$ for Order 2. From our spectroscopic fits we find that the longer wavelengths tend to exhibit small phase offsets and a range of phase curve relative amplitudes (see Fig.~\ref{fig:dobbs-dixon}). 

In contrast, at shorter wavelengths we observe an unexpected trend: phase offsets increasing to the east and retaining large relative amplitudes. The high relative amplitude at these wavelengths is due to the nightside emitting little to no flux, making our measured minimum flux effectively zero. This suggests that the day-night recirculation efficiency is still inefficient at these wavelengths, but the brightest location is shifted consistently eastward. At optical wavelengths, we might expect reflected light to impact the offset; however, reflected light measurements tend to favor a westward offset due to condensates evaporating on the hotter eastern hemisphere \citep{Parmentier2016,Coulombe2025-LTT-NIRISS-PC,Kennedy2025}. Despite the higher irradiation temperature of WASP-121\,b, clouds and refractory condensates can still form on the nightside and may be transported to the western limb \citep{Roman2021,Mikal-Evans2023,Komacek-ultrahotclouds}. Alternatively, the short-wavelength emission might probe different atmospheric depths where the circulation is baroclinic, producing altitude-dependent hot spot shifts \citep[e.g.,][]{Changeat2024}. In that case, the observed eastward offset would not represent a permanent equilibrium state.

The surprising result of large eastern offsets at short wavelengths is supported by our independent fits from \texttt{NAMELESS} (see Fig \ref{fig:Phase-offsets}). However, the exact values of the offsets and relative amplitudes depend on the fitting framework and on how negative planetary flux is treated. For example, the \texttt{exoTEDRF} fit shows a smaller phase offset than \texttt{NAMELESS} especially within Order 2, likely due to the inclusion of ellipsoidal variation and for Order 2, the stellar variability model. Excluding the stellar variability model brings the offsets from the two fits into closer agreement. Likewise, 

when the \texttt{exoTEDRF} fitting framework is applied to the \texttt{NAMELESS} reduction, the offsets and relative amplitudes become consistent across the two pipelines.

\subsection{Comparison to Bell\_EBM} \label{sec:EBM-Grid}

We compare our observations to a simple, semi-analytical energy balance model, \texttt{Bell\_EBM}, which accounts for the effects of H$_2$ dissociation \citep{Bell2018}. This process effectively increases the nightside temperature, thereby reducing the day–night temperature contrast. The model assumes there is a single, fully mixed atmospheric layer that absorbs all of the incident radiation and that flows zonally at a constant velocity. This toy model has been previously used to fit phase curve observations \citep[e.g.,][]{Dang2022, Davenport2025}. Comparing to the effective temperature, the EBM should always be fit to the bolometric light-curve.

We construct a precomputed grid of energy balance models tailored to the planet. The model includes the option to invert a bolometric light-curve into effective temperatures, allowing us to directly compare our observed effective temperatures to the predictions of \texttt{Bell\_EBM}. Our grid is generated by fixing the system parameters to the values from \cite{Sing2024-NIRSpec} with the exception of the orbital period and inclination that we take from our Order 1 white light-curve fit.

For each grid point, we vary three key model parameters: wind speed (v$_{\rm wind}$, in km/s), mixed layer depth ($P_0$, in bar), and Bond albedo ($A_\mathrm{B}$). The zonal wind speed, v$_{\rm wind}$, is assumed to be constant everywhere on the planet. The mixed layer pressure corresponds to the bottom of the mixed layer, the portion of the atmosphere that responds to the time-varying stellar flux experienced by a parcel of gas. We first ran a broad grid search before narrowing on the more likely values and creating a finer grid. The final grid consists of 11 albedo values, $ A_\mathrm{B}\in [0,0.35]$; 50 wind speeds v$_{\rm wind} \in [0,2]$\,km/s; and 100 mixed layer depths $P_0 \in [0.01,4.5]$\,bar for a total of 55,000 models. For each model, we compute the $\chi^2$ with respect to the observed phase variations. 

\subsection{Reflected and Thermal Components}

\begin{figure}[h]
    \centering
    \includegraphics[width=0.495\textwidth]{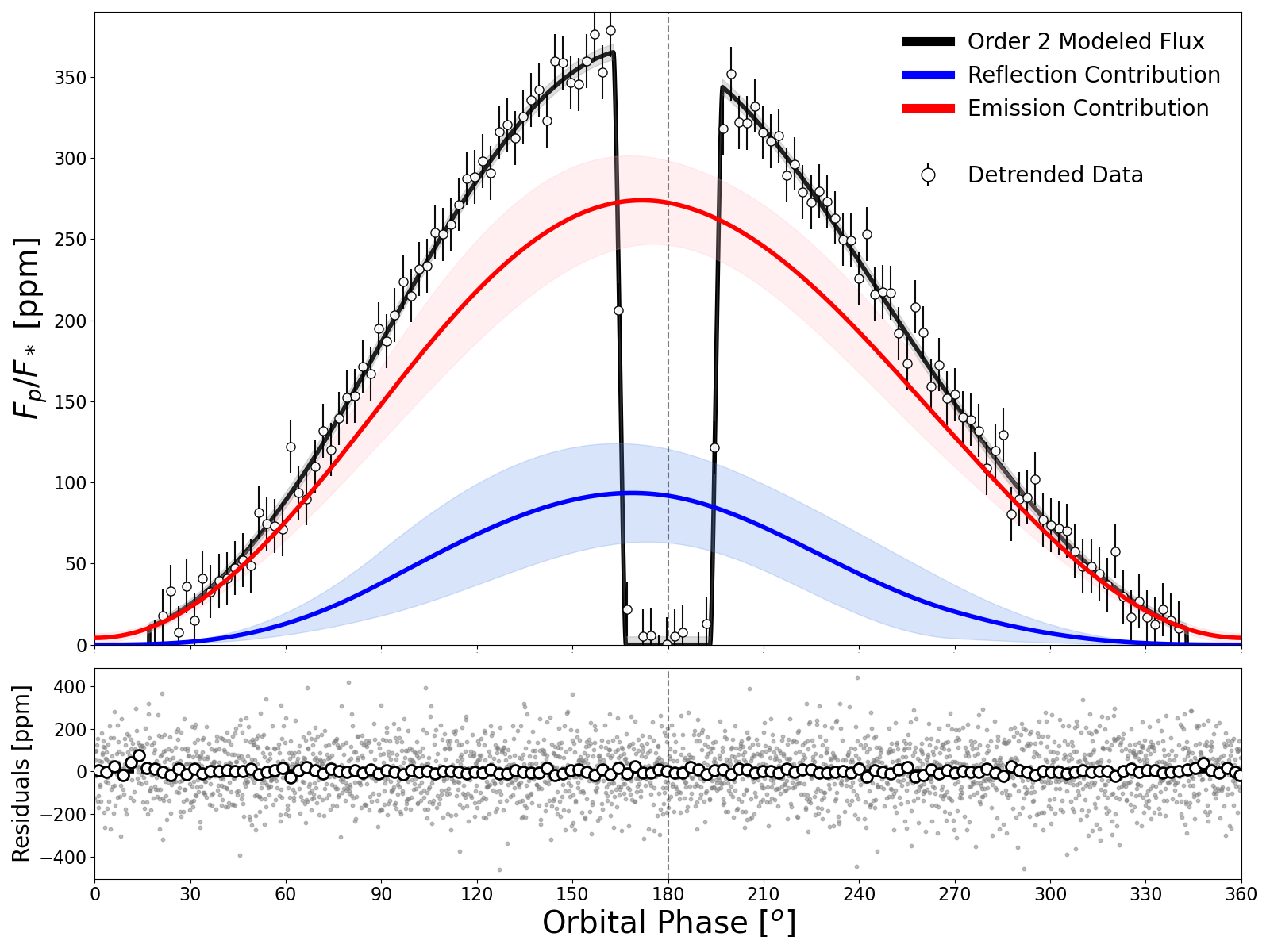}
    
	\caption{\emph{Top:} Reflected light and thermal emission components from the Order 2 detrended white light-curve. Model is described in Section \ref{sec:Slice-Model} and was fit four longitudinal slices across the planet; slices are centered at $\phi$ = 45$^\circ$, 135$^\circ$, 225$^\circ$, 315$^\circ$. Red shows thermal emission, blue shows reflected light and black shows the full model. Solid lines show the median retrieved value while shaded regions show the 1 $\sigma$ uncertainties on each model component. The detrended data are shown in black, binned with 20 points per bin for visual clarity. \emph{Bottom:} Residuals of median model to the data. Larger points are residuals binned to 20 points per bin.}
	\label{fig:Order2-slice_model}
\end{figure}

We present results from the Order 2 detrended white light-curve [0.6--0.85 $\mu$m, effective $\lambda \approx 0.72 \, \mu$m] to the slice model described in Section \ref{sec:Slice-Model}. Our fits, which account for both thermal and reflected light, indicate that thermal emission dominates the observed flux. In fact, thermal emission alone is sufficient to explain the data. While reflected light can contribute to the flux observed near the dayside, it cannot account for the nonzero flux around transit as the planet reflects virtually no light toward the observer at these phases. Nonetheless, we include results considering both thermal emission and reflected light (Figure \ref{fig:Order2-slice_model}) to properly marginalize over the full range of possibilities and constrain the amount of reflected light.

As shown in Figure \ref{fig:Order2-slice_model}, thermal emission and reflected light are highly degenerate at Order 2 wavelengths. The 1$\sigma$ confidence regions are derived by sampling each parameter over 2000 steps from the burned-in chain and estimating the standard deviation of each model component. 

We present the best-fit model using the median retrieved parameter values. From the retrieved albedo values we follow \citet{Schwartz2015} and infer the geometric albedo as
\begin{equation}
    A_g = \delta_{\rm refl}\bigg(\frac{a}{R_{\rm p}}\bigg)^2,
\end{equation}
\noindent where $\delta_{\rm refl}$ is the reflected light eclipse depth. We estimate the reflected light at eclipse as
\begin{equation}
    F_r(\phi_o=0) = \delta_{\rm refl} =   \sum^N_{i=1} \bigg[A_i \times\,K_{\rm reflect, i}(\phi_o = 0)\bigg].
\end{equation}
\noindent From this, we measure a geometric albedo of $A_\mathrm{g} = 0.093^{+ 0.029}_{- 0.027}$ or a 3$\sigma$ upper limit of 0.175. This is consistent with the geometric albedo of  $0.07^{+0.037}_{-0.040}$ found at a similar bandpass in TESS [0.6--0.95\,$\mu m$] \citep{Daylan2021}. Similarly, an eclipse in the z$^\prime$ band (central wavelength of 0.9\,$\mu m$) suggests a geometric albedo of $0.16 \pm 0.11$ \citep{Mallonn2019}. Nevertheless, our data provide evidence of reflected light in Order 2.

We note that when increasing the number of longitudinal slices (from four to six or eight), the model prefers higher levels of reflected light. Given this sensitivity, we elect to present the simplest model with four slices for consistency.

\begin{figure}[h]
    \centering
    \includegraphics[width=0.495\textwidth]{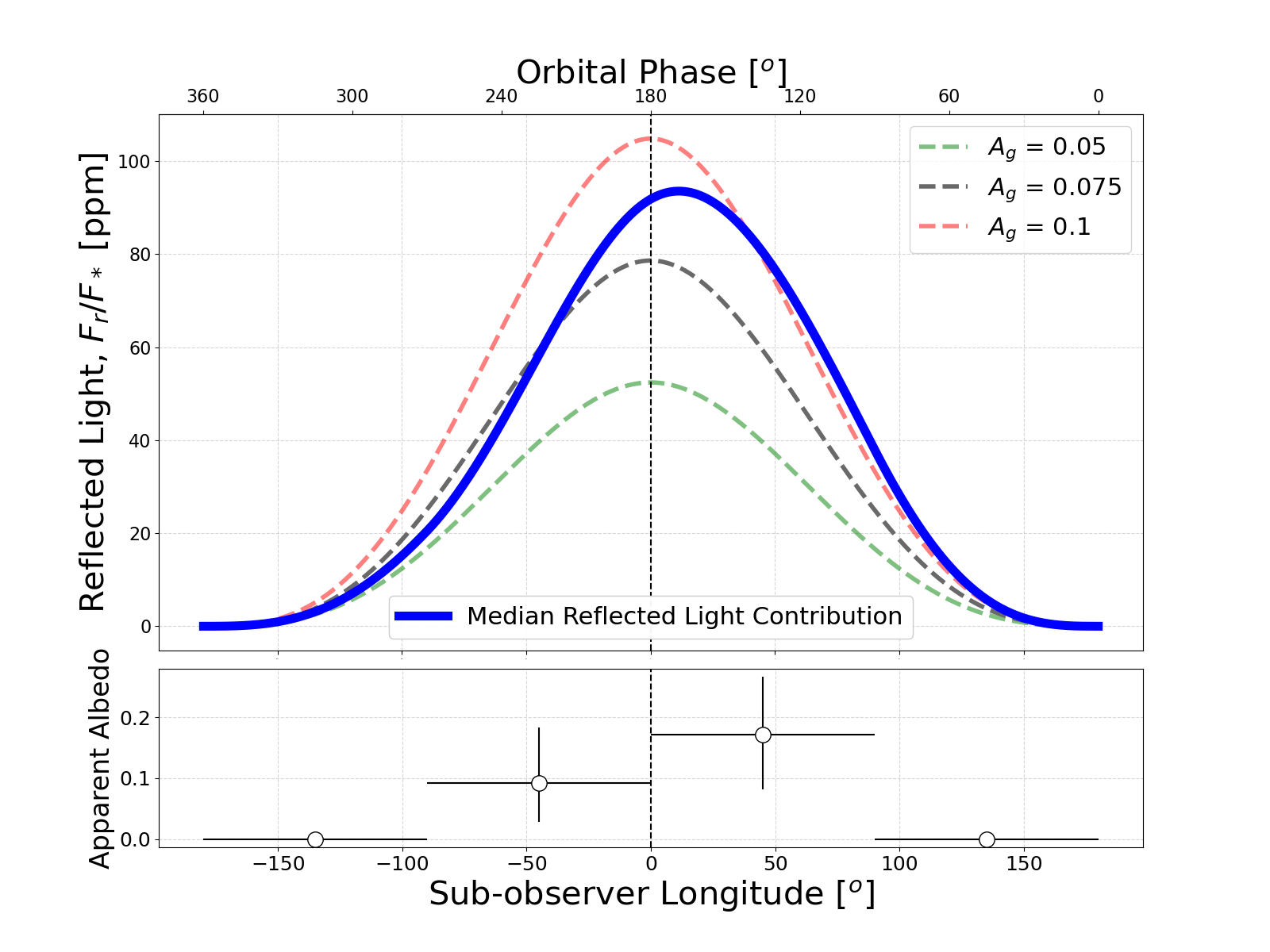}
    
	\caption{\emph{Top Panel:} Median reflected light contribution (blue) compared to Lambertian reflectors (dashed lines) with geometric albedos of 0.05 (green), 0.075 (gray) and 0.1 (red). Plotted in terms of the sub-observer longitude which is inversely related to the orbital phase (top ticks). \emph{Bottom Panel:} The measured apparent albedo for longitudinal slices centered at $\phi$ = 45$^\circ$, 135$^\circ$, 225$^\circ$, 315$^\circ$. The sub-observer longitudes are then shifted to be from -180$^\circ$ to +180$^\circ$ for better visible clarity. Observations show a small geometric albedo compared to the Bond albedo and perhaps an eastward reflected offset.}
	\label{fig:Order2-reflected-light}
\end{figure}

We compare the median reflected contribution to a Lambertian reflector in Figure \ref{fig:Order2-reflected-light}. The reflected light component shows a distribution that is asymmetric and shifted eastward, suggesting a slightly more reflective eastern dayside of the planet. This is in apparent contradiction with the expectation of inhomogeneous cloud cover, which results in greater reflectivity west of the substellar point \citep{Parmentier2016,Coulombe2025-LTT-NIRISS-PC}.

We tested our slice model using 4, 6, and 8 longitudinal slices, allowing albedo values to go negative with a lower prior limit of $A_i =-1$. One consistent trend emerges across all tests: eastward reflection is always favored. For slices closest to the substellar point, the eastern slice consistently exhibits a higher albedo than the western slice.

\section{Discussion} \label{sec:Discussion}
We presented a new analysis of the ultra-hot Jupiter WASP-121\,b from its JWST/NIRISS/SOSS phase curve. The wavelength range of this instrument captures the majority of the planet's bolometric flux, allowing for the most accurate energy budget measurement to date.

\subsection{Observational Systematics and Fit Discrepancies} 
To test the robustness of our results, we analyzed the data using two independent reduction and fitting pipelines. Our fiducial analysis, performed with \texttt{exoTEDRF}, incorporates physically motivated components such as ellipsoidal variations, a stellar variability model in the Order 2 wavelengths, and a prior punishing negative planetary flux. In contrast, the \texttt{NAMELESS} pipeline omits ellipsoidal variations and stellar variability, and allows negative planetary flux solutions with no constraints. The most significant differences between the two approaches emerge in the inferred nightside temperature and phase offsets, which we explore in detail in the following subsections. Importantly, when the \texttt{exoTEDRF} fitting framework is applied to the \texttt{NAMELESS} reduction, the results are consistent with our fiducial analysis, demonstrating that the observed differences arise from the modeling assumptions rather than the data reduction.

From our white light-curve fits, if the Gaussian process model described in Section \ref{sec:GaussianProcess} corresponds to stellar granulation, then the stellar granulation time scale is $30.05^{+4.44}_{-3.61}$ minutes for Order 1 and $35.5^{+6.20}_{-5.17}$ minutes for Order 2. The corresponding stellar granulation amplitudes are $83^{+5.2}_{-4.9}$ ppm (Order 1) and $95^{+6.7}_{-6.0}$ ppm (Order 2),  which are consistent with the expected amplitude of 96 ppm \citep{Gilliland2011}. However, the measured granulation time scales are significantly longer than the expected value of 6 minutes. A similar discrepancy was reported by \cite{Coulombe2025-LTT-NIRISS-PC}.

Increasing the stellar effective temperature from $T_* = 6459 \pm 140 \, K$ \citep{Delrez2016-discovery} to $T_* = 6628 \pm 66 \, K$ \citep{Sing2024-NIRSpec} has a slight impact on the calculation of the brightness temperatures, which in turn increases our estimates for the planet's effective temperature by approximately 30\,K. This has a minimal impact on estimations of the planet's day and nightside temperatures. 
The more important effect is for the planet's irradiation temperature. In the analytic model of \cite{Cowan2011-albedo}, increasing the stellar temperature shifts the Bond albedo to higher values for a given planet/star contrast (Equation\,\ref{equation:albedo-recirc}). Therefore, we update the Bond albedo and heat recirculation from previous phase curve observations using the new inferred irradiation temperature. 

There are multiple sources of uncertainty for the Bond albedo and heat recirculation efficiency. For our analysis of WASP-121\,b, the smallest contribution comes from $a/R_*$, with relative uncertainties of $\sim$ 0.01\%. A larger uncertainty of $\sim$ 1\% stems from the stellar effective temperature. The measured eclipse depths are also a significant source of uncertainty, with relative uncertainties of $\sim$ 1\% for Order 1 and $\sim$ 5\% for Order 2. Since Bond albedo and heat recirculation efficiency depend on the planet's bolometric flux, the greatest uncertainty in any inference ultimately comes from the fraction of captured total flux by a given observation. NIRISS is the first instrument to capture more than half of the planetary spectral energy distribution.

\subsection{Atmospheric Energy Transport}

The dayside and nightside temperature measurements in different bandpasses produce different inferred Bond albedos and heat recirculation efficiencies. However, even our two independent analyses of the same dataset do not overlap at their 3$\sigma$ confidence regions. The \texttt{exoTEDRF} fit gives $A_\mathrm{B} = 0.277\, \pm0.016$ and $\epsilon = 0.246\, \pm 0.014$, while the \texttt{NAMELESS} fit yields $A_\mathrm{B} = 0.307 \pm 0.018$ and $\epsilon = 0.186^{+0.021}_{-0.019}$ (Fig.~\ref{fig:albedo-recirculation}).

This discrepancy is driven almost entirely by differences in the inferred nightside effective temperature. In the \texttt{exoTEDRF} fit, the penalty on negative planetary flux allows a broader range of wavelengths to contribute to the nightside temperature estimate, resulting in a narrower and systematically warmer distribution. In contrast, \texttt{NAMELESS} permits negative planetary flux values with no penalties, which are excluded from contributing during the Monte Carlo estimation of the effective temperature. As a result, fewer data points inform the nightside temperature, broadening the distribution and biasing it toward cooler values (see Fig. \ref{fig:Nightside-Brightness-Temp}). Although both fits return consistent dayside temperatures, the $\sim$120\,K offset in nightside temperature leads to significantly different Bond albedo and heat recirculation efficiency estimates. These results underscore how methodological choices can dominate the interpretation of a planet’s atmospheric energy transport, underscoring the critical role of nightside flux modeling in phase‑curve analyses.

\begin{figure}[htb]
    \centering
    \includegraphics[width=0.495\textwidth]{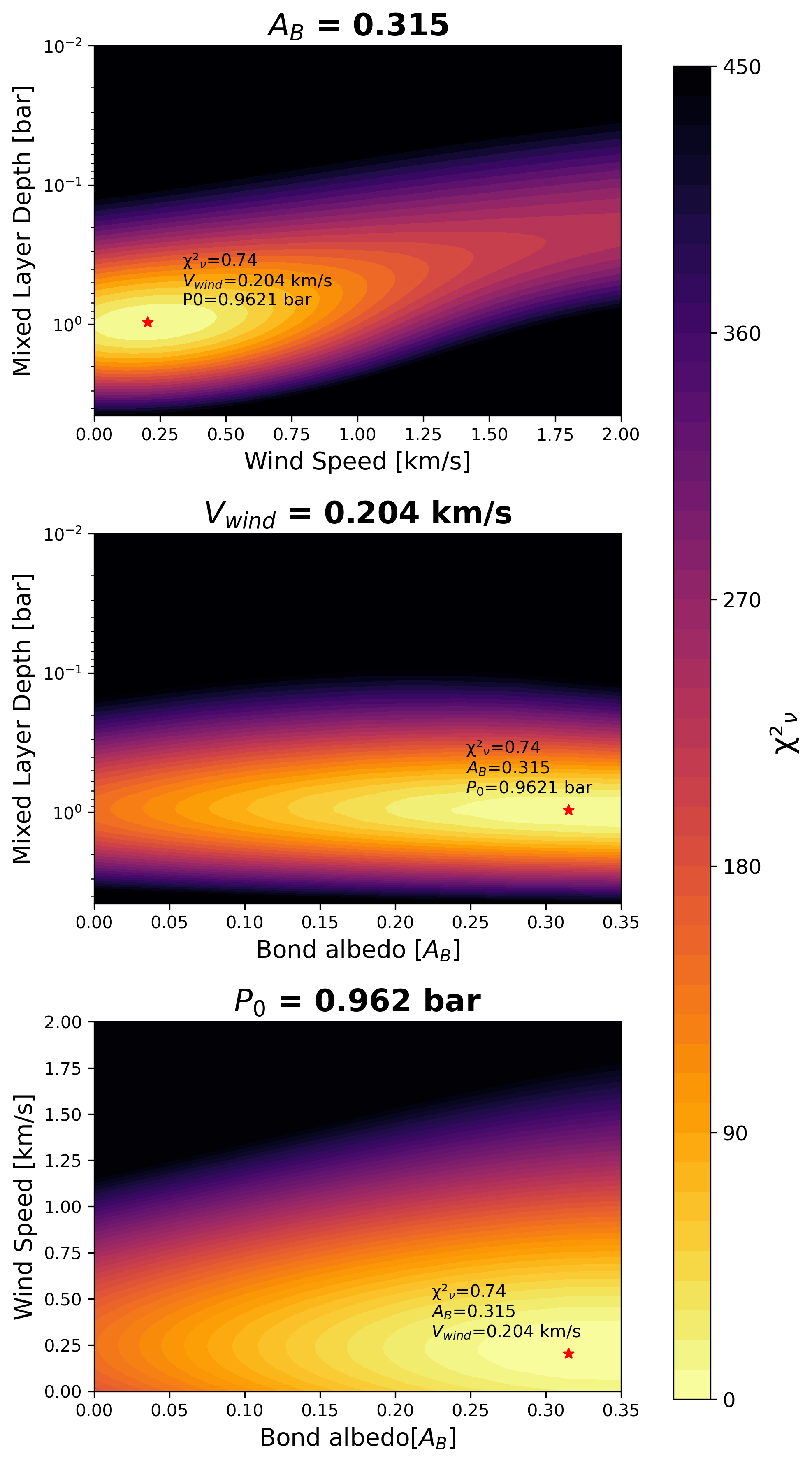}
    
	\caption{Contour plots from the \texttt{BELL\_EBM} grid showing parameter slices from the best matched model to the planetary effective temperature. A red star is plotted in each contour to guide the eye to the location of the best matched model. Bond albedo is in line with simpler estimates based solely on dayside and nightside spectra, $P_0$ is roughly in line with expected deposition depth of incoming stellar radiation, and estimated wind speed is much slower than expected.}
    
	\label{fig:EBM-Grid}
\end{figure}

Comparing the phase variations in effective temperature to a grid of \texttt{BELL\_EBM} energy balance models, we find the best models in the grid prefer a Bond albedo of $\sim$0.31, along with low wind speeds of $\sim$0.20\,km/s and a mixed layer depth near 1\,bar (see Figure \ref{fig:EBM-Grid}. This suggests the EBM favors slightly higher Bond albedos than those inferred solely from dayside and nightside temperatures. However, the Bond albedos remain broadly consistent between the two analyses. The energy balance models also prefer a mixed layer depth pressure of $\sim$1\,bar: higher regions of the atmosphere respond to the time-variable stellar flux, while deeper layers do not. This is consistent with theoretical expectations for hot Jupiter atmosphere's photon deposition layer near 1\,bar \citep{Parmentier2013, Tan2019}. This result contrasts with \texttt{BELL\_EBM} fits to WASP-121\,b’s Spitzer 3.6 and 4.5 $\mu$m phase curves \citep{Davenport2025}, which inferred mixed layer depths of less than 0.2 bar. The deeper mixed layer found in our analysis is more consistent with expectations for hot Jupiters, though differences in wavelength coverage and observational constraints likely contribute to these variations.

Our \texttt{BELL\_EBM} fits favor wind speeds of approximately 200\,m/s. This is larger than the sub-50 m/s values found in \texttt{BELL\_EBM} fits to WASP-121\,b’s Spitzer 3.6 and 4.5 $\mu$m phase curves \citep{Davenport2025} but slower than the global zonally averaged wind speeds of 400–500 m/s predicted by general circulation models \citep{Davenport2025}. However, our inferred wind speed is not directly comparable to equatorial jet speeds measured in previous studies \citep{Sing2024-NIRSpec, Seidel2025}. First, \texttt{BELL\_EBM} assumes a single global wind speed throughout the atmosphere, whereas high-resolution Doppler spectroscopy has revealed a complex vertical structure in WASP-121\,b’s atmospheric circulation. \cite{Seidel2025} measured wind speeds of 4–10 km/s from neutral iron lines tracing lower pressures ($<10^{-2}$ bar). If our inferred mixed layer depth of $\sim$1 bar is accurate, then lower wind speeds at these pressures are plausible, with much faster jet streams and day-to-night winds confined to the upper atmosphere. Additionally, JWST NIRSpec observations reveal a wavelength-dependent radial velocity amplitude between NRS1 and NRS2, indicative of average zonal winds of $\sim$5.2 km/s. However, wind speeds inferred from \texttt{BELL\_EBM} are more analogous to the planet’s bolometric energy transport, likely probing deeper layers than these high-altitude tracers. The inferred slow wind speeds could be a consequence of magnetic drag, in which the planet’s magnetic field impedes atmospheric flow and reduces heat transport efficiency in WASP-121\,b \citep{Perna2010, Menou2012}.

Hotter planets are generally expected to have shorter radiative timescales resulting in larger day-night temperature contrasts and, therefore, higher phase curve amplitudes \citep{Cowan2011-albedo,Komacek&Showman2016}. However, observations to date do not consistently show this trend \citep{Dobbs-Dixon&Cowan2017, Crossfield2015}, possibly because phase curves often captured only a fraction of a planet’s total bolometric flux. 
This apparent discrepancy has motivated further theoretical work exploring other factors that may influence phase curve behavior. For instance, ultra-hot Jupiters typically show smaller phase offsets due to their short radiative timescales \citep{Cowan&Agol-timescales}. Additional processes—such as magnetic drag \citep{Beltz2022, Beltz2024} and nightside clouds \citep{Komacek-ultrahotclouds,Parmentier2021}—can further shape atmospheric dynamics, enhancing amplitudes and reducing offsets. Conversely, H$_2$ dissociation and recombination may act to redistribute heat more efficiently across the planet, potentially reducing amplitude and increasing the offset \citep{Bell2018,Tan2024}.

Another factor to consider is the comparison between different observed bandpasses. General circulation models (GCMs), which incorporate more complex physics and atmospheric mixing, generally offer more realistic predictions than simpler semi-analytic models. However, GCMs without atmospheric drag tend to overpredict phase offsets for WASP-121\,b \citep[e.g.,][]{Parmentier2018}. Recent observations and GCM studies now suggest that WASP-121\,b is indeed experiencing atmospheric drag \citep{Wardenier2024}.

Phase offsets of WASP-121\,b have been all over the map. Measurements with TESS [0.6--0.95 $\mu$m] for WASP-121\,b show a slight westward offset of $-6.3 \pm 3.9^\circ$, consistent with zero\citep{Daylan2021}. Analysis of two Hubble WFC3 phase curves [1.12--1.64 $\mu$m] reported a broadband eastward offset of $6.4 ^{+1.7 \, \circ}_{-2.3}$ with spectroscopic offsets spanning 0--10$^\circ$ \citep{Mikal-Evans2022}. A later re-analysis of the same phase curves using a three-region model (and therefore not directly comparable to phase offsets) indicated a shift in hot spot position between the two epochs and favored models with a $\sim$30$^\circ$ eastward displacement and a hot spot extent of $\sim$50$^\circ$ \citep{Changeat2024}. The first analysis of the NIRSpec white light-curves reported slight eastward offsets of $3.4 \pm 0.11^\circ$ with NRS1 [2.70--3.72 $\mu$m] and $2.7\pm 0.12^\circ$ with NRS2 [3.82--5.15 $\mu$m] \citep{Mikal-Evans2023}. A more recent full analysis confirms these small eastward shifts in the white light-curves, reporting $2.95^{+0.11}_{-0.12}$$^\circ$ with NRS1 and $2.39^{+0.12}_{-0.13}$$^\circ$ with NRS2, and further extends the results to spectroscopic phase curves, which likewise show slight eastward offsets but never exceeding $10^\circ$ \citep{Evans-Soma2025-NIRSpec-PhaseCurve}. However, analysis of the Spitzer phase curve in similar bandpasses to NRS1 and NRS2 have reported westward offsets. The analysis of \citet{Morello2023} find offsets of  $-5.9 \pm 1.6^\circ$ ($3.6 \, \mu$m) and $-5.0^{+3.4\circ}_{-3.1\circ}$ ($4.5 \, \mu$m), while a reanalysis of $4.5 \, \mu$m  Spitzer phase curves found an offset of $-5 \pm 4^\circ$ for WASP-121\,b \citep{Dang2025}. Furthermore, another independent re-analysis of both Spitzer phase curves of WASP-121\,b finds offsets consistent with zero, measuring $-0.8 \pm 1.9^\circ$ at $3.6 \, \mu$m and $0.42 \pm 1.74^\circ$ at $4.5 \, \mu$m \citep{Davenport2025}.

From our spectroscopic fits we find that the longer wavelengths tend to exhibit small phase offsets (eastward offset of 5.1 $\pm$ 1.4$^\circ$ in Order 1 white light-curve) and a range of phase curve relative amplitudes showing less stark day/night temperature contrasts suggesting potential evidence that we are probing different layers of the atmosphere with varying dynamical timescales (see Fig.~\ref{fig:dobbs-dixon}).

\subsection{Reflected Light and Albedo Constraints}

If the geometric albedo of WASP-121\,b is comparable to the inferred Bond albedo of 20--30\%, we would expect reflected light to dominate the planetary flux at the shortest wavelengths of NIRISS/SOSS. Assuming a gray albedo, this transition from reflected light to thermal emission dominance occurs within the Order 2 wavelengths. Fitting for the thermal emission and reflected light components simultaneously is the only way to properly estimate the contribution of both. Indeed, as shown earlier, fitting the detrended Order 2 white light-curve for both a thermal and reflection component reveals that reflection likely has a non-negligible impact. 
Regardless of the number of longitudinal slices fit or whether we allow albedo to be negative, we consistently find that reflected light peaks on the eastern side of the planet. 

The peak in reflection on the eastern side of the planet may be influenced by its strong correlation with thermal emission. However, if this eastward reflection is real, it could be further supported by the detection of large phase offsets at the shortest wavelengths. These offsets appear robust across different modeling approaches (Figure \ref{fig:Phase-offsets}), with the \texttt{NAMELESS} fit showing even larger offsets. A notable discrepancy emerges between the two spectral orders, with Order 2 exhibiting a sudden shift to lower phase offsets compared to Order 1. This is likely driven by the inclusion of a stellar variability model, which peaks before secondary eclipse (see Figure \ref{fig:WLC-fits}). When the stellar variability model is excluded, the phase offsets between the two orders become more consistent. Despite these differences, the \texttt{exoTEDRF} fits still show consistently large phase offsets at short wavelengths, a striking and unexpected result. We further note that the trend to higher eastern offsets appears as high as $\sim$ 1.4\,$\mu$m, indicating this is likely astrophysical rather than a systematic difference between Order 1 and Order 2. GCM simulations of this planet, including those that account for clouds, do not find a trend of increasingly eastward offsets at shorter wavelength (Frazier et al., Under Review). 

It has been known for a decade that hot Jupiters exhibit greater Bond albedos than geometric albedos, the so-called albedo paradox \citep{Schwartz2015,Crossfield2015}. The Bond albedo is constrained through thermal emission, while the geometric albedo is typically determined from reflected light at short optical wavelengths. In this work, we constrain both albedos within a single observation by probing the planet’s reflection and emission and across optical and near-infrared wavelengths. Due to the unprecedented captured flux of our measurement, we provide the most precise estimate of Bond albedo for an exoplanet to date. Our results indicate a relatively high Bond albedo of 20-–30\% but a low geometric albedo $\sim$10\%, a factor of 3 discrepancy. This finding further sharpens the albedo paradox, underscoring the need for future studies to reconcile the reflected light observations of WASP-121\,b and other hot Jupiters to better understand the underlying atmospheric processes.

A possible resolution to the albedo paradox lies in the wavelength dependence of reflected light. The geometric albedo we have measured has been constrained mainly from near-infrared wavelengths [0.6--0.85 $\mu$m] which represents only 23\% of the reflected stellar flux, as stars emit most of their light at shorter wavelengths (as for Fig \ref{fig:Captured-Flux-Fraction}, but this time for the stellar effective temperature). Assuming a two-component approximation for the Bond albedo, $A_B = f_{\rm long}A_{\rm red} + f_{\rm short}A_{\rm blue}$, where $f$ denotes the fractional stellar flux in each band and $A$ the corresponding albedo, we can estimate the implied reflectivity at shorter wavelengths, assuming $f_{\rm red} = 0.23$. If we adopt our measured near-infrared geometric albedo ($A_{\rm long} \approx 0.09$) and a Bond albedo of 0.27, this would require $A_{\rm short} \approx 0.34$—suggesting significantly enhanced reflectivity at bluer wavelengths. Future observations that constrain reflected light at shorter optical wavelengths will be essential to test this scenario and determine whether the albedo paradox stems from incomplete spectral coverage.

\section*{Acknowledgments}

This project was undertaken with the financial support of the Canadian Space Agency through a NEAT/GTO grant. The contributions from RF were supported in part by Grant \#2019-1403 from the Heising-Simons Foundation. N.B.C. acknowledges support from an NSERC Discovery Grant, a Tier 2 Canada Research Chair, and an Arthur B.\ McDonald Fellowship and thanks the Trottier Space Institute and l'Institut de recherche sur les exoplanètes for their financial support and dynamic intellectual environment.  L.D. is a Banting and Trottier Postdoctoral Fellow and acknowledges support from the Natural Sciences and Engineering Research Council (NSERC) and the Trottier Family Foundation. S.P. acknowledges support from the Swiss National Science Foundation under grant 51NF40\_205606 within the framework of the National Centre of Competence in Research PlanetS. R.J.M. is supported by NASA through the NASA Hubble Fellowship grant HST-HF2-51513.001, awarded by the Space Telescope Science Institute, which is operated by the Association of Universities for Research in Astronomy, Inc., for NASA, under contract NAS 5-26555. R.A. acknowledges the Swiss National Science Foundation (SNSF) support under the Post-Doc Mobility grant P500PT\_222212 and the support of the Institut Trottier de Recherche sur les Exoplanètes (IREx). D.L acknowledges financial support from NSERC and FRQNT. C.P.-G. acknowledges support from the E. Margaret Burbidge Prize Postdoctoral Fellowship from the Brinson Foundation.  JDT acknowledges funding support by the TESS Guest Investigator Program G06165. Some/all of the data presented in this article were obtained from the Mikulski Archive for Space Telescopes (MAST) at the Space Telescope Science Institute. The specific observations analyzed can be accessed via \dataset[doi: 10.17909/ppjp-9743]{https://doi.org/10.17909/ppjp-9743}.


\software{
    \texttt{astropy} \citep{astropy, astropy2, astropy3},
    \texttt{batman} \citep{batman}, 
    \texttt{celerite} \citep{Foreman-Mackey2017-celerite}, 
    \texttt{emcee} \citep{emcee}, 
    \texttt{exoTEDRF} \citep{exotedrf, Radica2023-AwesomeSOSS, Feinstein2023-WASP39b}, 
    \texttt{jwst} \citep{jwst}, 
    \texttt{matplotlib} \citep{matplotlib}, 
    \texttt{numpy} \citep{numpy}, 
    \texttt{pandas} \citep{pandas}, 
    \texttt{scipy} \citep{scipy}
}

\begin{appendices}
\onecolumngrid
\setcounter{figure}{0}
\renewcommand{\thefigure}{C\arabic{figure}}
\setcounter{table}{0}
\renewcommand{\thetable}{C\arabic{table}}

\appendix

\section{NAMELESS data reduction and light-curve fitting} \label{app:nameless}

An additional reduction of the data is produced using the NAMELESS data reduction pipeline \citep{Coulombe2023,Coulombe2025-LTT-NIRISS-PC} following the methodology described in \cite{Allart2025}. In short, we go through all Stage 1 and Stage 2 steps of the STScI \texttt{jwst} pipeline until the flatfield correction step. We then correct for bad pixels by interpolating over pixels whose mixed spatial second derivative are $>4\sigma$ away from the local median. The non-uniform background is subtracted from all integrations by scaling uniformly the two portions of the model background provided by STScI that are separated by the sharp jump in background flux near $x=700$. Cosmic rays are corrected for by computing the running median of all pixels and bringing any count that is a $>3\sigma$ outlier to the value of its pixel's running median. We correct 1/$f$ noise by scaling individually each column of the order 1 and 2 traces and computing the constant additive that best reduces the chi-square (considering only pixels that are $<30$ pixels away from the center of the trace), which is then subtracted from the data. Finally, we extract the stellar spectra using a fixed box aperture of width 40 pixels. The order 1 and 2 white light-curves produced from this reduction are shown in Figure \ref{fig:exoTEDRF-NAMELESS}.

Two sets of spectroscopic light-curves are produced by binning the data at fixed resolving powers of $R=100$ and $R=300$. The full fitting methodology is presented in \cite{Pelletier2025-NIRISS-emission} and is described briefly below. The astrophysical model consists of a second order sinusoid ($\sum_{i=0}^{N=2}F_n\cos(n[\varphi(t)-\delta_n])$, where $\varphi(t)$ is the orbital phase) which is combined to a transit and eclipse model computed using \texttt{batman} \citep{batman}. We adopt large uniform priors for the coefficients of the sinusoidal model, allowing for negative flux. The transit parameters consist of the planet-to-star radius ratio ($R_\mathrm{p}/R_\star$) and the quadratic limb-darkening coefficients [$u_1$, $u_2$] ($\mathcal{U}[-3,3]$, following \citet{Coulombe2024}). For the systematics model, we consider a linear slope in time and a Heaviside step function centered at the time where the tilt event occurs ($S(t)=c+v(t-t_0)+j\Theta(t-t_\mathrm{tilt})$). We also freely fit the photometric scatter for each spectral bin. The spectroscopic fits are ran assuming zero-eccentricity ($e=0$), a mid-transit time of $T_0=60244.520384$\,MJD, an orbital period of $P=1.2749250$\,days \citep{Patel2022}, as well as values of $a/R_\star=3.754$ and $b=0.15$ for the scaled semi-major axis and impact parameter as measured from the NAMELESS order 1 and 2 white-light-curves. The light-curve fits are ran with the \texttt{emcee} Python package \citep{emcee} for 6250 steps considering 4 walkers per free parameter. All free parameters are fit considering wide uniform priors. The resulting day- and nightside spectra for the $R=300$ fit are shown in Figure \ref{fig:Day-v-nightside} and the phase curve offset spectrum from the $R=100$ fit is shown in Figure \ref{fig:Phase-offsets}.

\section{Analytical Solutions to Thermal Emission and Reflected Light Kernels}

Here, we present analytical solutions to the thermal emission and reflected light kernels presented in section \ref{sec:Slice-Model} in equations \ref{eq:reflect_kernel} and \ref{eq:emission_kernel}.

\subsection{Thermal Emission Kernel}

Given that $d\Omega = \sin{\theta} d\theta d\phi$ and writing out the visibility function (equation \ref{eq: vis}), the integral for thermal emission is:
\begin{align}
  K_{{\rm thermal},\, i} &= \frac{1}{\pi} \oint V(\theta,\phi, t)\, d\Omega \\
                         &= \frac{1}{\pi} \int_{\phi_{i}}^{\phi_{i+1}} \int_{0}^{\pi} 
                            \left[ \sin^2\theta \sin\theta_o \cos(\phi - \phi_o) 
                            + \sin\theta \cos\theta \cos\theta_o \right]
                            \, d\theta \, d\phi.
\end{align}

\noindent The final simplified solution is:
\begin{equation}
 \Rightarrow K_{{\rm thermal},\, i} = \frac{\sin\theta_o}{2}\sin(\phi - \phi_o)\Biggr|_{\phi_i}^{\phi_{i+1}}
\end{equation}

This solution comes with a couple of caveats in order to reach physical solutions. The solution is always valid when $\phi_{i}$ and $\phi_{i+1}$ are both on the visible hemisphere, but when one or the other is on the far side of the planet, the respective slice edge needs to replaced by $\pi/2$  or $-\pi/2$.

\subsection{Reflected Light Kernel}

The process for the reflected light kernel is similar, with the addition of extra normalization factor and the illumination function (equation \ref{eq: illumin}). This integral then becomes:


\begin{align}
    K_{{\rm reflect},\, i} 
    &= F_*(\lambda) \left(\frac{R_*}{a} \right)^2 \frac{1}{\pi} \int^{\phi_{i+1}}_{\phi_i} V(\theta,\phi, t) I(\theta, \phi, t) \, d\Omega \\
    &= \int_{\phi_i }^{\phi_{i+1}} \int_{0}^{\pi} 
    \Big[\left(\sin\theta \sin\theta_o \cos(\phi - \phi_o) + \cos\theta \cos\theta_o\right) \notag \\
    &\quad \quad \qquad \qquad \times \left(\sin\theta \sin\theta_s \cos(\phi - \phi_s) + \cos\theta \cos\theta_s\right) 
    \sin\theta \Big] \, d\theta \, d\phi \,.
\end{align}

\noindent The complete solved kernel is then: 
\begin{align}
 \Rightarrow K_{{\rm reflect},\, i} 
    &=\left( \frac{2}{3\pi} \right) F_*\left( \frac{R_*}{a} \right)^2 \Bigg[\sin\theta_s \sin\theta_o \bigg(\cos(\phi_o - \phi_s) \, \cdot\phi 
 + \frac{\sin(2\phi - \phi_o - \phi_s)}{2}\bigg)+\cos\theta_s \cos\theta_o\, \phi \Bigg]_{\phi_i }^{\phi_{i+1}}.
\end{align}

\noindent Simplifying with $\phi_s = 0$ and dividing by the stellar flux we are left with:

\begin{align}
\frac{F_{\rm reflect}}{F_*}=\left( \frac{2}{3\pi} \right) \left( \frac{R_p}{a} \right)^2 \Bigg[ \cos(\phi_o)(\phi_{i+1} - \phi_i) +
\frac{1}{2} \left( \sin(2\phi_{i+1} - \phi_o) - \sin(2\phi_i - \phi_o) \right) \Bigg].
\end{align}

This solution also carries the same caveats that the thermal emission kernel solution had to remain physical. Namely, the slice edges must be within the visible hemisphere, $-\pi/2 < \phi_o<\pi/2$. 

\section{Additional Plots}
\begin{figure*}[htb]
    \centering

    \includegraphics[width=\linewidth]{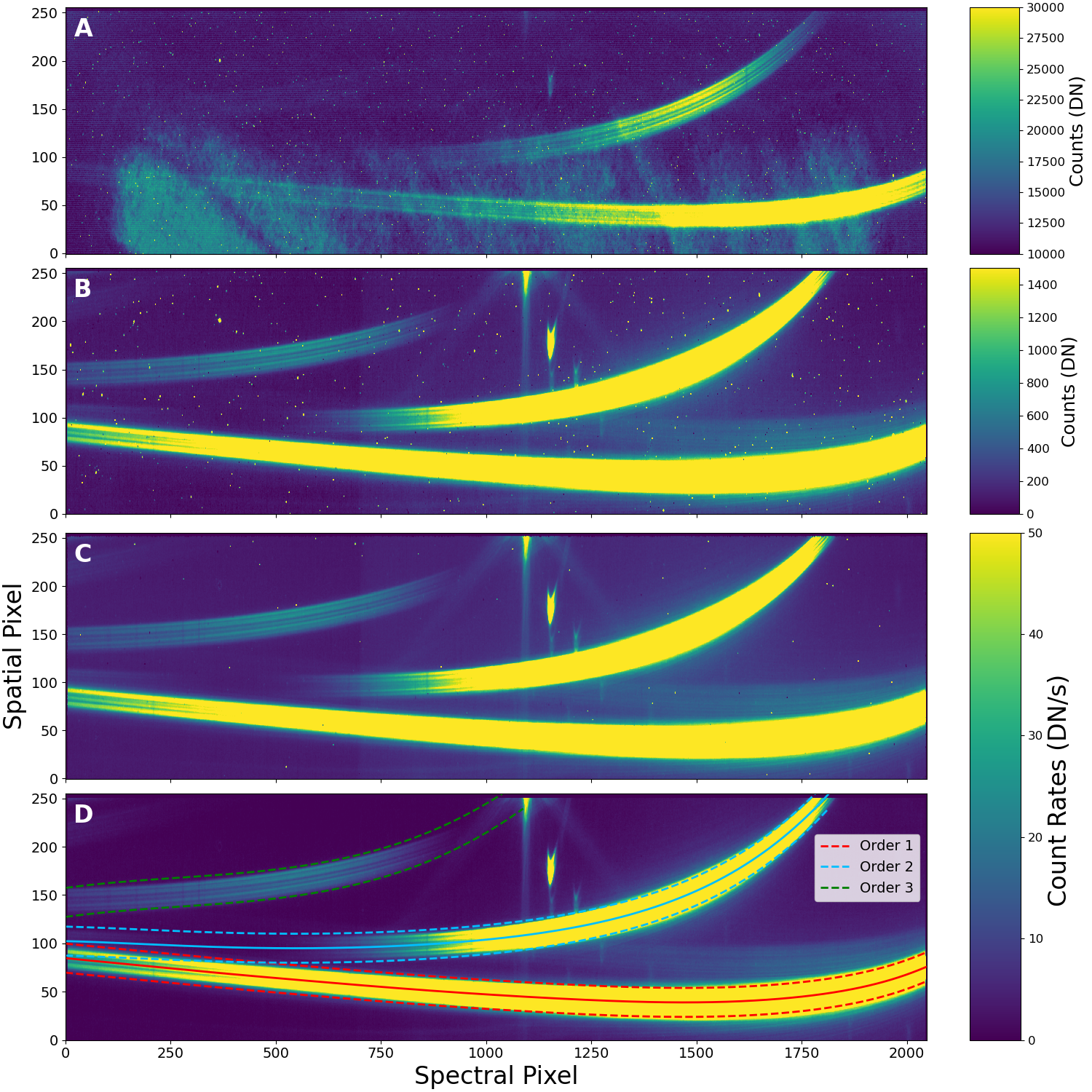}
	\caption{Visualization of \texttt{exoTEDRF} reduction steps. \textbf{(A)}  A raw, uncalibrated frame in counts of data numbers (DN). \textbf{(B)} Frame after superbias subtraction and $1/f$ correction on group level. \textbf{(C)} Frame after ramp fitting, at the end of Stage 1 -- detector level processing operations.  \textbf{(D)} Final data product at the end of Stage 2 -- spectroscopic processing operations including second background subtraction, flat-field correcting and bad pixel correction. The centroids for the spectral orders including the extraction width of 30 pixels are included in red (Order 1), blue (Order 2) and green (Order 3). Only Order 1 and 2 are used in this analysis. There are two bright order 0 contaminants located around x = [1000,1150] and y = [125,250]. With the absence of a GR700XD/F277W exposure for this observation, we are unable to measure the dilution effects of the contaminants within the spectral trace.}
	\label{fig:reduction-steps}
\end{figure*}

\begin{figure*}[htb]
    \centering
    \includegraphics[width=0.95\textwidth]{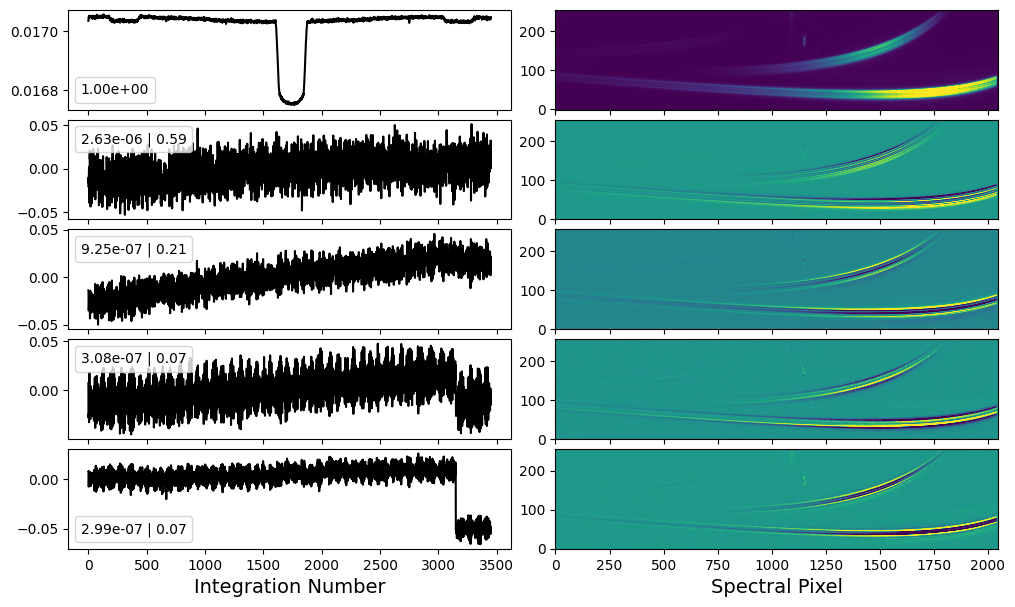}
    
	\caption{Diagnostic plot showing results on five component PCA of time series to explain variance in the data. Left column corresponds to the eigenvalues while right column shows the corresponding eigenimages. First component shows the white light-curve while other components correspond to detector-based signals (thermal beating pattern, position drifts, etc.) and noise. Components 4 and 5 revealed the presence of a minor tilt event. We use PCA components 2--4 in our light-curve fitting (see Methods).}
	\label{fig:PCA-plot}
\end{figure*}

\begin{figure*}[htb]
    \centering
    \includegraphics[width=0.95\textwidth]{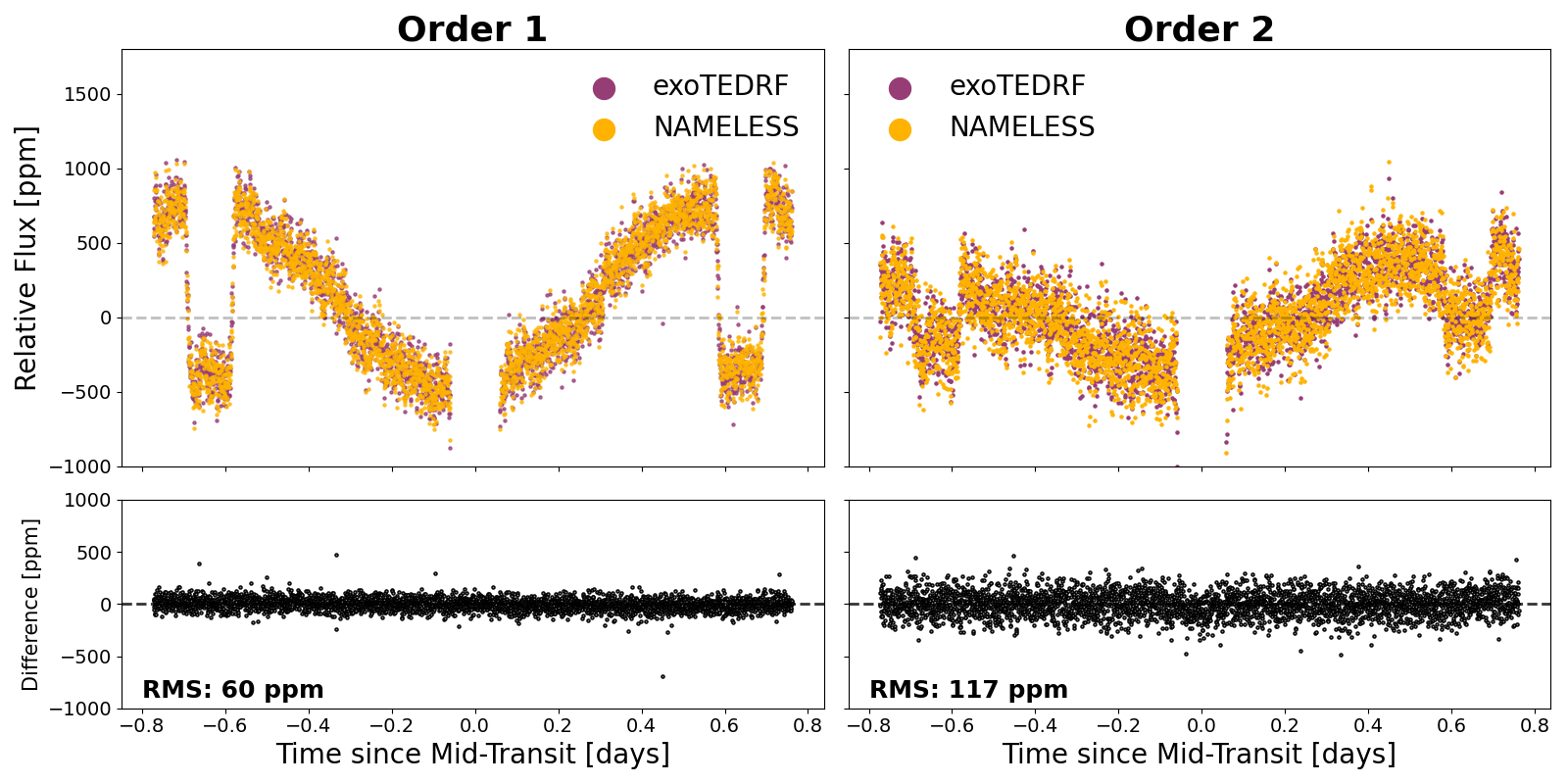}
    
	\caption{Comparison of reduction pipelines between \texttt{exoTEDRF} (purple) and \texttt{NAMELESS} (gold) white light-curves for Order 1 (left) and Order 2 (right). A difference between the two reductions (\texttt{exoTEDRF} - \texttt{NAMELESS}) is shown in the bottom row with the RMS of the differences in the lower left corner. The two reductions agree well and show no noticeable differences.}
	\label{fig:exoTEDRF-NAMELESS}
\end{figure*}

\begin{figure*}[h]
    \centering
    \includegraphics[width=0.95\textwidth]{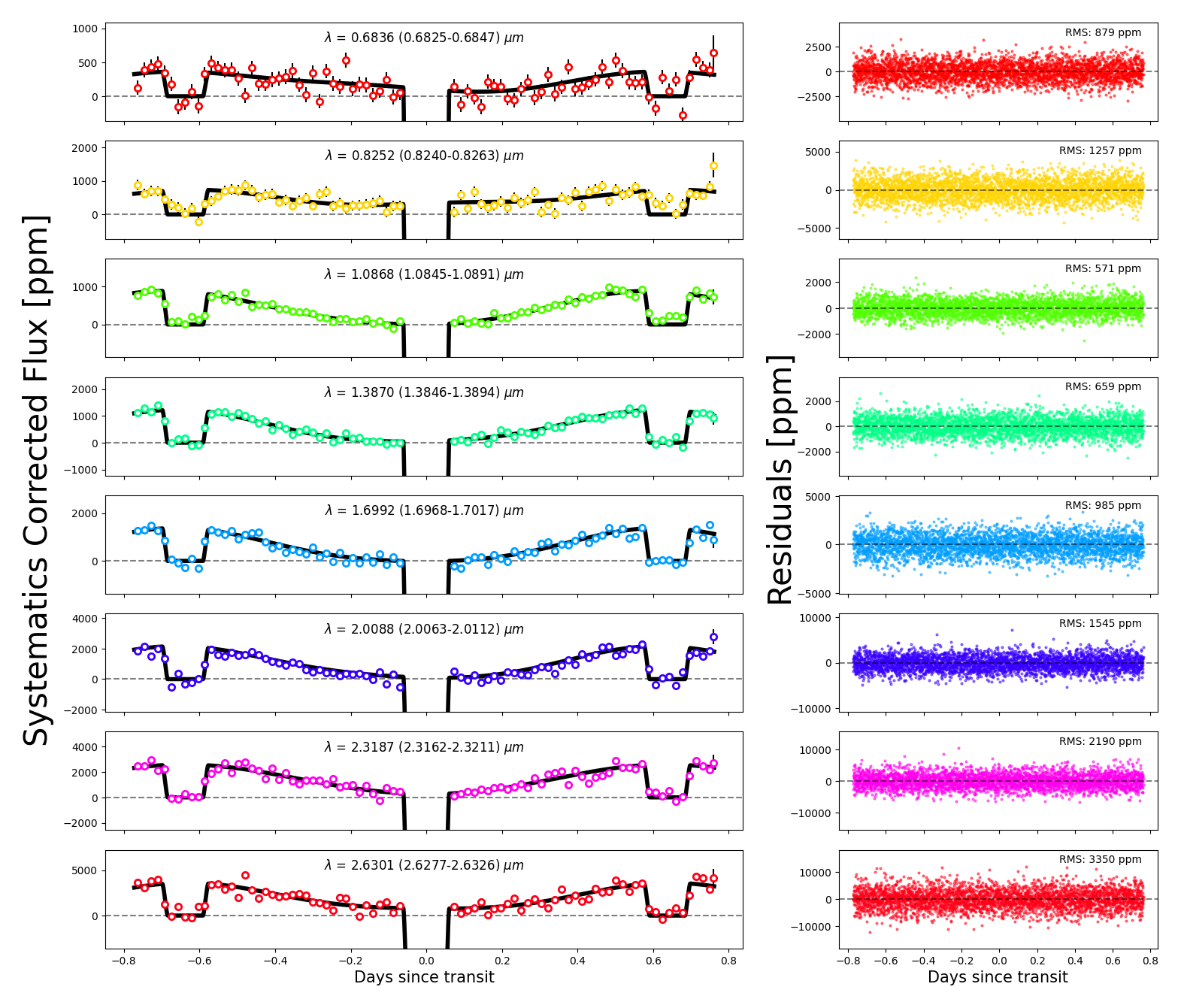}
    
	\caption{Fits of systematics corrected spectroscopic light-curves at various wavelengths at the 5 pixel per bin level. First 2 light-curves are from Order 2 and the rest from Order 1 to show full wavelength coverage. \emph{Left:} Spectroscopic light-curves (colored points) with the best fit astrophysical model (black line). The data are fitted considering all integrations but are shown with forty integrations per bin for visual clarity. \emph{Right:} Residuals of the best fit astrophysical model to the detrended data shown at all integrations. A black dashed line is shown at zero to guide the eye. The root mean square of the residuals is shown in the top right.}
	\label{fig:spectro-light-curves}
\end{figure*}

\begin{figure}[htb]
    \centering
    \includegraphics[width=\linewidth]{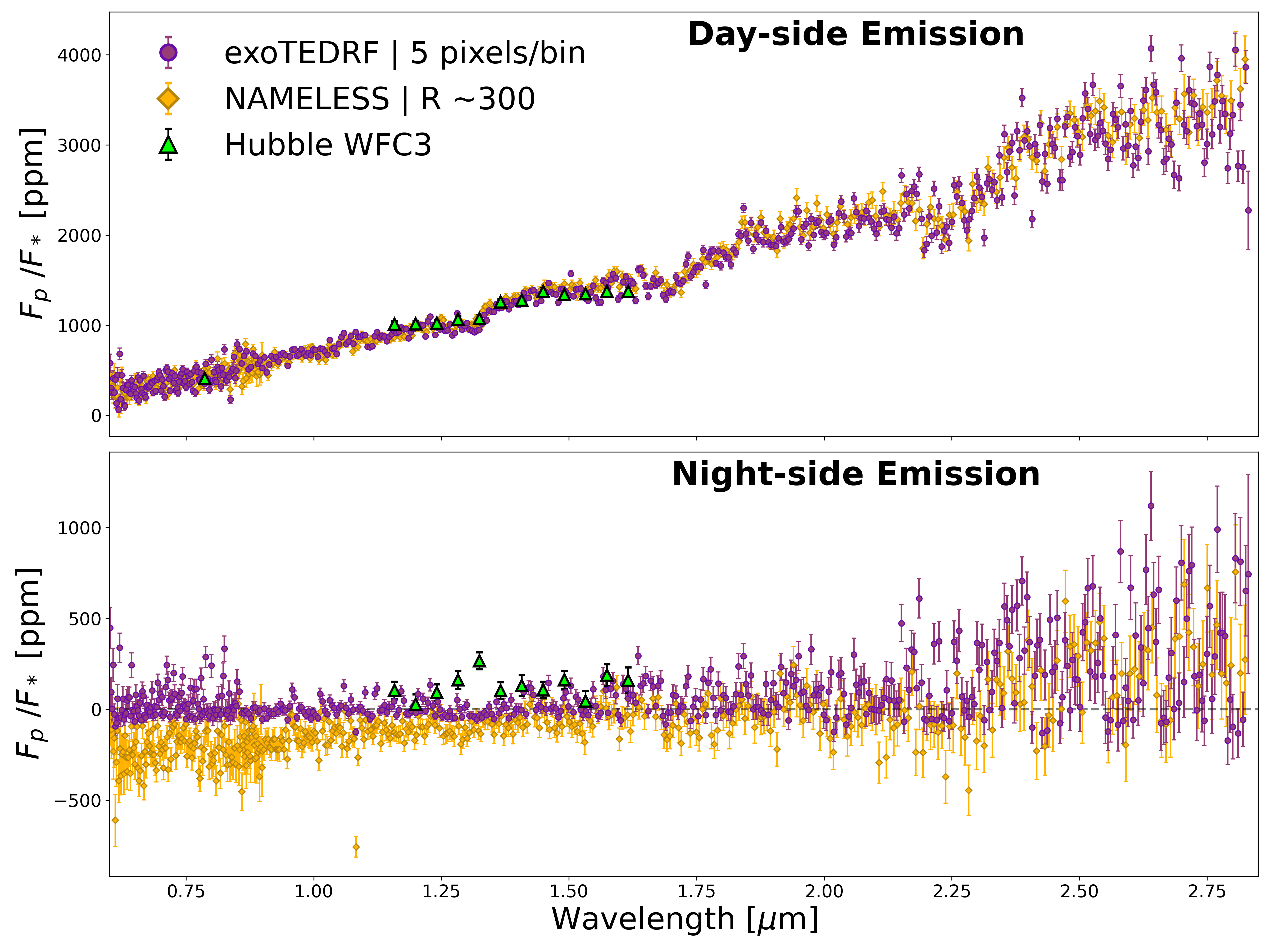}
    
	\caption{Nightside and dayside emission spectra from \texttt{exoTEDRF} (purple) and \texttt{NAMELESS} (gold) fits. As discussed in section \ref{sec:reduction-fit-differences}, the \texttt{exoTEDRF} fit penalizes negative planetary flux while \texttt{NAMELESS} does not, accounting for the differences in the nightside emission. Both model assumptions have implications on the inferred nightside temperatures as seen in Figure \ref{fig:Nightside-Brightness-Temp}.}
	\label{fig:Day-v-nightside}
\end{figure}

\begin{figure}[htb]
    \centering
    \includegraphics[width=\linewidth]{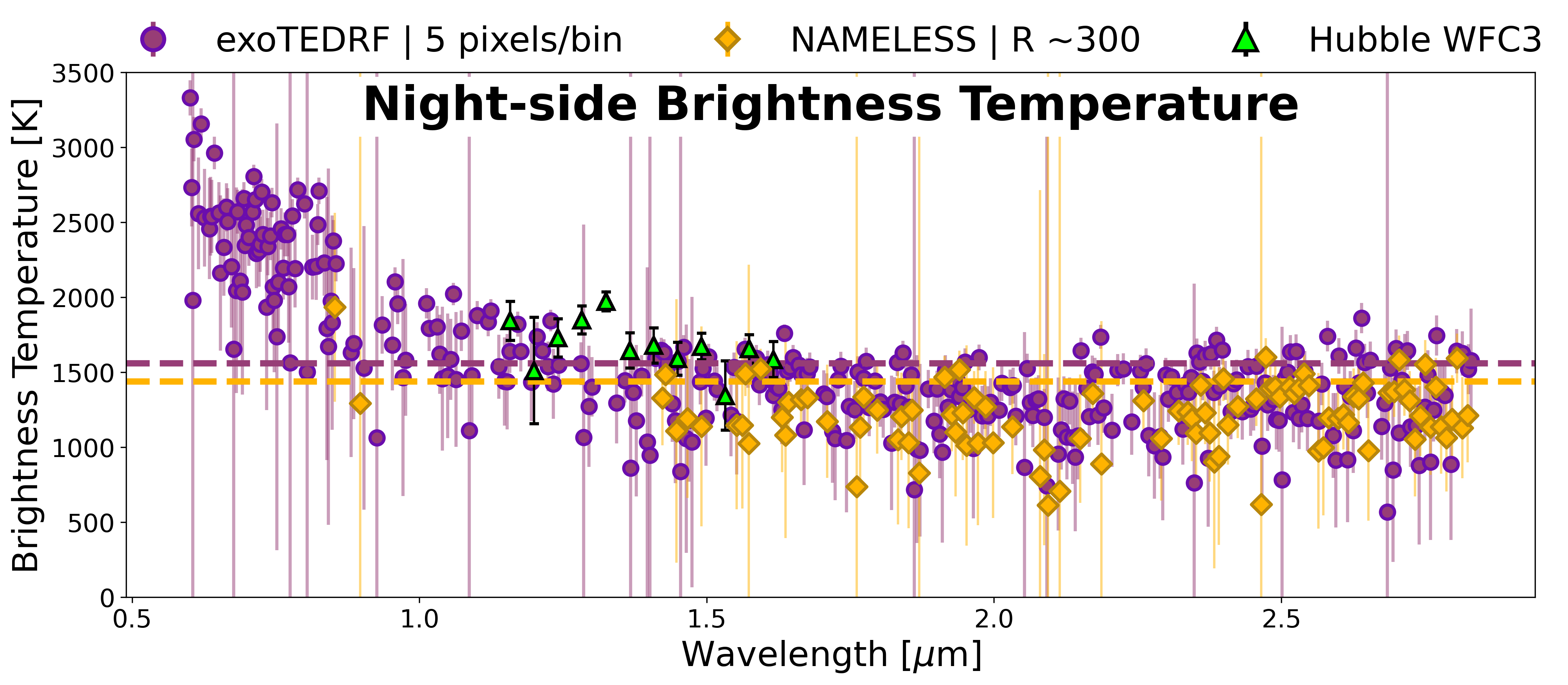}
    
	\caption{Nightside brightness temperature from \texttt{exoTEDRF} (purple) and \texttt{NAMELESS} (gold) nightside emission seen in the bottom panel of Figure \ref{fig:Day-v-nightside}. Negative flux values cannot be converted into a brightness temperature thereby the \texttt{exoTEDRF} fit has more points in brightness temperature at shorter wavelengths. Dashed vertical lines show the nightside effective temperature for \texttt{exoTEDRF} (purple) and \texttt{NAMELESS} (gold) showing that the inclusion of more brightness temperatures moves the effective temperature slightly higher. }
	\label{fig:Nightside-Brightness-Temp}
\end{figure}

\begin{figure}[h]
    \centering
    \includegraphics[width=\linewidth]{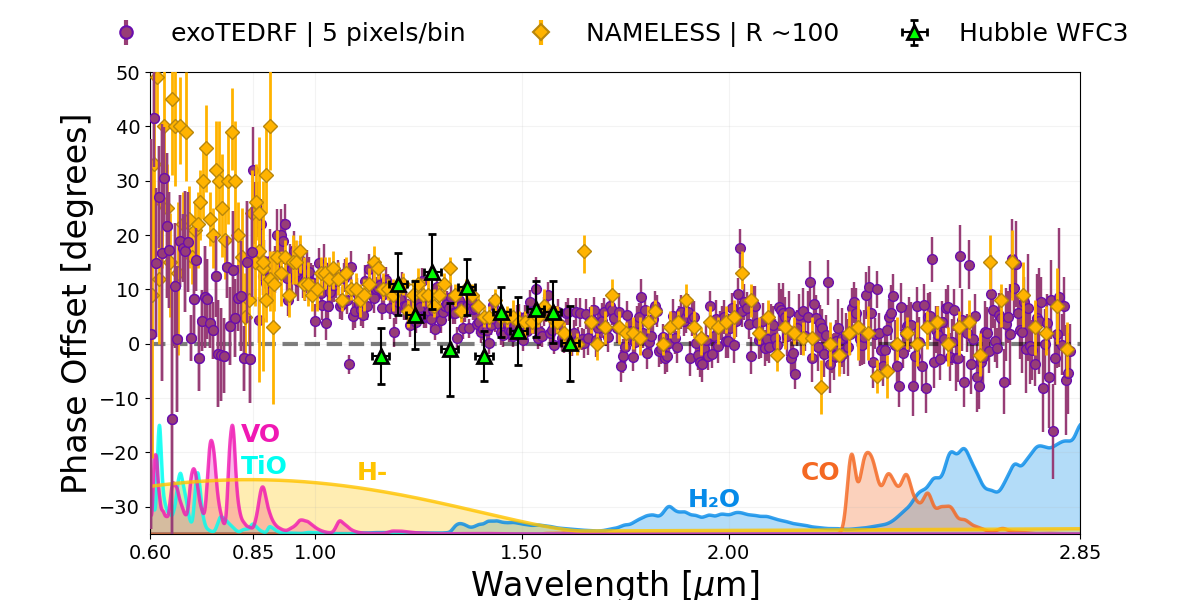}
    
	\caption{Phase offsets from the \texttt{exoTEDRF} fit (purple) and \texttt{NAMELESS} fit (gold). Colors at the bottom of plot show normalized and scaled cross-sections for TiO (cyan), VO (pink), H$^-$ (yellow), H$_2$O (blue) and CO (orange) in order to visualize possible contributions to the phase offsets. Only H$^-$ could appear to trace the increase at shorter wavelengths. The larger offsets in the \texttt{NAMELESS} fit likely result from the inclusion of ellipsoidal variation in \texttt{exoTEDRF}, except in Order 2, where the addition of a stellar variability model causes a sharp drop in phase offsets. For clarity, Order 2 phase offsets from \texttt{exoTEDRF} are binned with three points per bin. Both reductions consistently show increasing eastward offsets at shorter wavelengths.}
	\label{fig:Phase-offsets}
\end{figure}

\begin{figure}[htb]
    \centering
    \includegraphics[width=\linewidth]{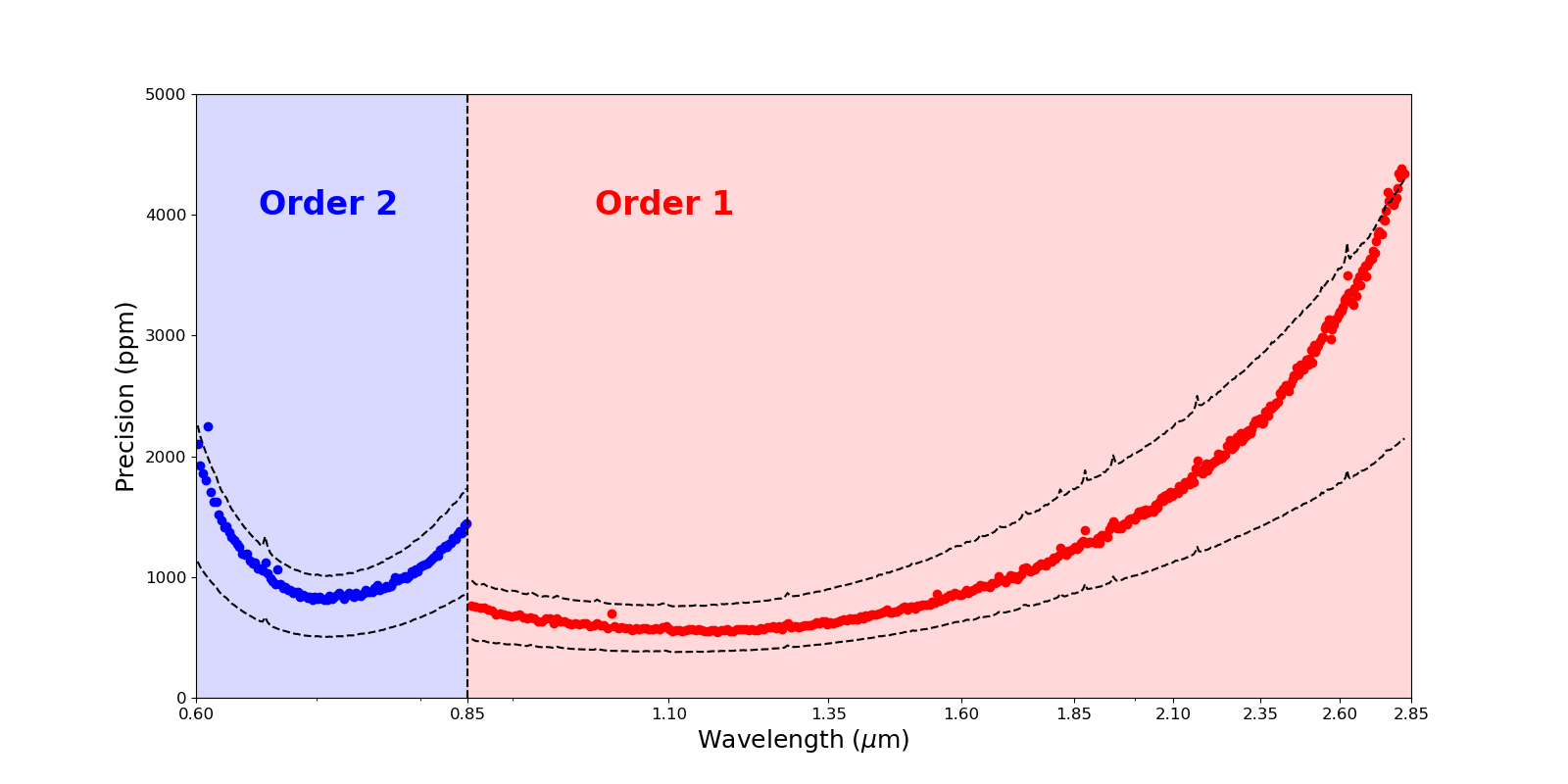}
    
	\caption{Spectroscopic precision from \texttt{exoTEDRF} fits of WASP-121\,b using 5pixels/bin. Dashed black lines show the photon noise limit and 2x the photon noise limit, respectively. Precision is generally 50\% above the photon limit with the exception of the shortest wavelengths in Order 2 and the longest wavelengths in Order 1.}
	\label{fig:error-spectrum}
\end{figure}

\begin{figure}[]
    \centering
    \includegraphics[width=\linewidth]{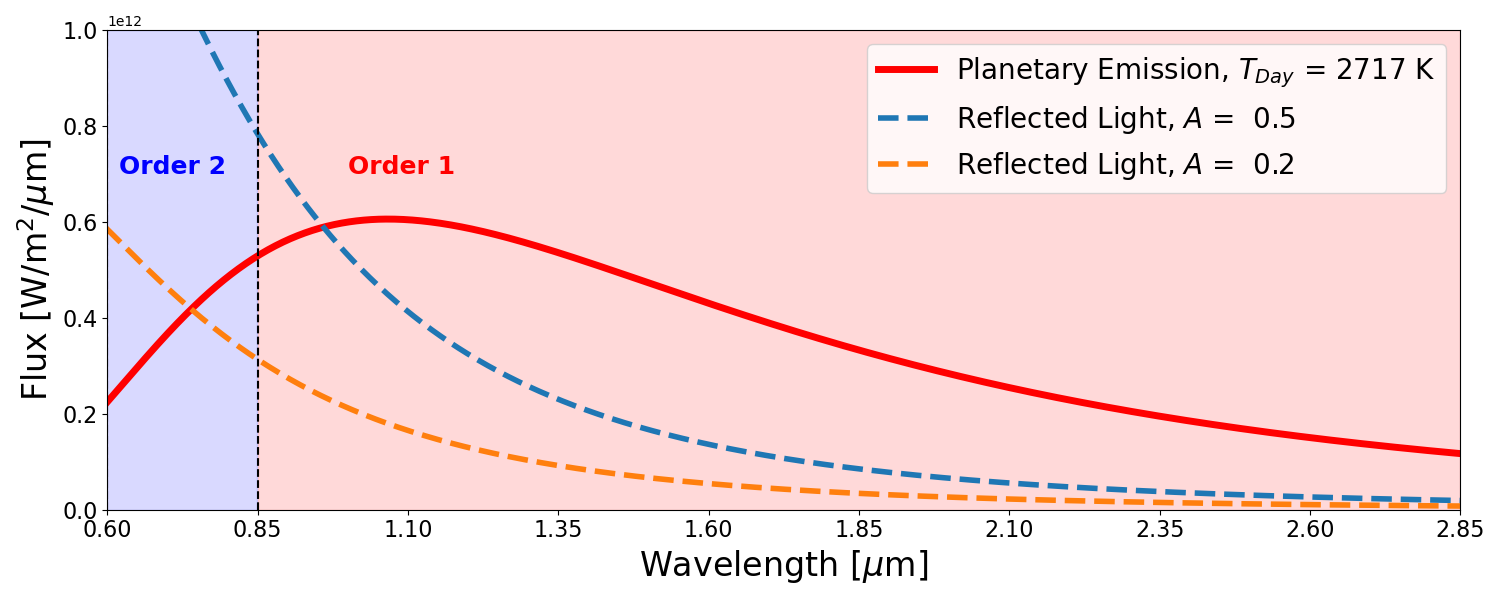}
    
	\caption{The level of contribution of reflected light and thermal emission coming from the planet if we assume the planet and star as a blackbody and a gray albedo. The red line shows the thermal emission from the planet given it's dayside $T_{\rm eff}$ value. The blue, orange and green dashed lines show the reflected light contribution for albedo values of 1, 0.5 and 0.2 respectively. Blue region shows the wavelength range covered by Order 2 while the red region shows for Order 1. With a bond albedo of 20-30\% we would expect reflected light to dominate over thermal emission at the shortest wavelengths of Order 2.}
	\label{fig:Estimated-reflected-light}
\end{figure}

Figures \ref{fig:reduction-steps}--\ref{fig:Estimated-reflected-light} provide additional visualizations and checks
that complement the results discussed in the main sections.

\twocolumngrid

\end{appendices}

\clearpage
\bibliography{references}
\bibliographystyle{aasjournal}
\end{document}